%% file: Sat_Comm_IoTJ.tex
\documentclass[10pt,journal,compsoc]{IEEEtran}

\usepackage{tikz}
\usetikzlibrary{shapes.geometric} 
\usetikzlibrary{arrows.meta}
\usetikzlibrary{calc,fit}

\ifCLASSOPTIONcompsoc
  \usepackage[nocompress]{cite}
\else
  \usepackage{cite}
\fi

\ifCLASSINFOpdf
 
\else
 
\fi

\usepackage{amsmath}
\usepackage{amsfonts}

\usepackage{url}
\usepackage[normalem]{ulem}
\usepackage{soul,color}

\soulregister\gls7
\soulregister\cite7
\soulregister\cref7
\soulregister\pageref7
\usepackage[acronym]{glossaries}
\renewcommand{\glossarysection}[2][]{}

\DeclareMathOperator*{\argmin}{arg\,min}
\usepackage[capitalise,noabbrev]{cleveref}
\usepackage{ragged2e}
\newglossary[slg]{symbolslist}{syi}{syg}{List of Symbols}

\loadglsentries[symbolslist]{symbols}
\include{glossary}
\include{acronyms}
\makeglossaries

\begin{document}
\newcommand{\figurewidth}{0.22}

\title{{Reinforcement Learning for Security Aware Computation Offloading in Satellite Networks}}

\author{Saurav~Sthapit, Subhash~Lakshminarayana, Ligang~He, Gregory~Epiphaniou and       Carsten~Maple~\IEEEmembership{}
\IEEEcompsocitemizethanks{\IEEEcompsocthanksitem S. Sthapit, G.~Epiphaniou and C. Maple are with the Warwick Manufacturing Group, University of Warwick, UK\protect\\
\IEEEcompsocthanksitem S. Lakshminarayana is with the School of Engineering, University of Warwick, UK\protect\\
\IEEEcompsocthanksitem L. He is with the Department of Computer Science, University of Warwick, UK\protect\\

E-mail: \{ saurav.sthapit, subhash.lakshminarayana, ligang.he, gregory.epiphaniou, cm \} @warwick.ac.uk
}
\thanks{}}

\glsresetall
\IEEEtitleabstractindextext{%
\begin{abstract}
The rise of \gls{ns} provides a platform for small and medium businesses to commercially launch and operate satellites in space. In contrast to traditional satellites, \gls{ns} provides the opportunity for delivering computing platforms in space. However, computational resources within space are usually expensive and satellites may not be able to compute all computational tasks locally. \acrfull{co}, a popular practice in Edge/Fog computing, could prove effective in saving energy and time in this resource-limited space ecosystem. However, \gls{co} alters the threat and risk profile of the system. In this paper we analyse security issues in space systems and propose a security-aware algorithm for \gls{co}. Our method is based on the reinforcement learning  technique, \acrfull{ddpg}.  We show, using Monte-Carlo simulations, that our algorithm is effective under a variety of environment and network conditions and provide novel insights into the challenge of optimised location of computation.\end{abstract}

}

\maketitle

\IEEEdisplaynontitleabstractindextext

\IEEEpeerreviewmaketitle

\IEEEraisesectionheading{\section{Introduction}\label{sec:introduction}}

\IEEEPARstart{T}he space industry is experiencing rapid growth currently, thanks to lowering technological and economic barriers to entry. \gls{cots} hardware such as Nvidia Jetson \footnote{\url{https://developer.nvidia.com/embedded/jetson-developer-kits}}{ and Xilinx Virtex FPGA} are readily available along with the plethora of software and support {\cite{xilinxRTKintexUltraScale2020,ludtkeOBCNGReconfigurableOnboard2014,pignolCOTSbasedApplicationsSpace2010}}. Similarly, \gls{sdn}\cite{sonCloudSimSDNNFVModeling2019,xuSoftwareDefinedNextGenerationSatellite2018} is continuing to revolutionise the way we connect, making it easier, more flexible and cheaper. Advances such as these have led to new commercial companies, including relatively \glspl{sme}, entering a space industry that has previously been restricted to large non-commercial organisations such as \gls{nasa}. This new environment has been coined as the `\gls{ns}' \cite{garymartinNewSpaceEmergingCommercial}. The \gls{ns} ecosystem comprises of thousands of satellites of all sizes and contrary to traditional satellites, Cubesats\cite{woellertCubesatsCosteffectiveScience2011} can be as small as $10 \times 10 \times 10~cm^3$. In this new paradigm, instead of acting solitary, the satellites may form a cluster or a constellation, communicate with each other, and jointly serve the users on the earth surface.  

In terms of applications, satellites used to be limited to relaying information from one point to another. Such architecture is commonly referred as bent-pipe architecture \cite{denbyOrbitalEdgeComputing2020}. However, modern satellites are intelligent; instead of being simple relays, they can sense, process and act intelligently \cite{akyildizInternetSpaceThings2019}. For example, a satellite can autonomously collect space debris or dock itself without human intervention\cite{mapleSecurityMindedVerificationSpace2020}. Satellites are also able to continuously monitor the environment using multiple sensors. In such cases, it is desirable to process the raw data in the orbit itself rather than transferring all of the data to Earth \cite{giuffridaCloudScoutDeepNeural2020}. 
However, this extra computation will add to the existing sensing and processing of the sensor data and not all of the satellites may be able to handle them \cite{furanoUseArtificialIntelligence2020} due to limitations in energy and computational power. 

Satellites, such as Cubesats will have to rely on other nearby satellites or space stations for processing their sensor data. Attempts have already been made to address these challenges. Recently, super computer satellites as small as kitchen microwaves are being launched in space \cite{ukspaceagencyTakeoffUKbuiltSupercomputer}. The objective of such super computer satellites is to offer `computing as a service' to other satellites in order to process the sensor data while in orbit. This process of offloading computation is already common in terrestrial computations for edge devices and is commonly referred to as \acrfull{co}.{ While \gls{co} in space is similar to \gls{co} on Earth in many respects, there exist some unique challenges. These include, (1) the server in space may not be as powerful as the server on Earth. This implies that there is a non-trivial queuing and computation delay at the server, which is not present in terrestrial applications. (2) The satellite network is very dynamic, especially in the \gls{leo} orbit. Hence, the topology of the satellite network will be changing rapidly.} (3) \gls{co} in space requires data to be transmitted to a different platform{ (satellite)}. This additional communication requirement will raise security risks, such as eavesdropping (from nearby satellites), data modification, and/or preventing the offloading satellite from accessing such service. Decisions regarding whether to offload and the level of security measures to be used in exchanging the data (between the server and the client satellite) are not trivial decisions. Careful consideration of the environment is required to assess if such offloading is beneficial in terms of time, energy, and the security risks incurred.

In this work, we explore \gls{co} in the context of satellites and \gls{ns} with the awareness of security threats in space. We formulate the security-aware \gls{co} problem as a multi-objective optimisation problem and jointly minimise the time, energy and security cost of the system using a \gls{rl} framework. Since our formulation involves decision variables that are continuous, we use the \gls{ddpg} method to solve the RL problem, since it can be directly applied to continuous action spaces, and avoids the need for discretisation \cite{Lillicrap2016ContinuousCW}. Our results show that even in the presence of wireless communications security threats, it is possible to offload computation and increase the efficiency of the system. The main contributions of the paper are as follows:
\begin{itemize}
	\item A new examination of the space landscape for communications security.
	\item The formulation of the security-aware \gls{co} problem within New space as a multi-objective problem.
	\item The development of a new \gls{ddpg}-based solution to solve the problem and analysis of its efficacy in comparison to an state of the art \gls{dqn} based solution.
\end{itemize}

We note that while there is extensive literature in wireless communications related to resource allocation/scheduling \cite{Sadr2009, SongRA2014, LakshWiOpt2011}, security \cite{ZouSecurity2016, LakshWiSec2018}, computational offloading \cite{shiEdgeComputingVision2016,maoSurveyMobileEdge24,abbasMobileEdgeComputing2018,machMobileEdgeComputing23,sthapitDistributedComputationalLoad2017,minLearningBasedComputationOffloading2019}, etc., the focus of all these works is on terrestrial mobile networks. In contrast, our work considers inter-satellite communications while incorporating the aforementioned domain-specific features. To the best of our knowledge, this work is the first to consider security-aware offloading in a satellite environment, and this is one of our important contributions.

The paper is structured as follows. \cref{sec:sat_arch} describes the  satellite architecture, inter-satellite communication and {various security risks} in space applications. In \cref{sec:co}, we define the basics of \gls{co} including the local execution and remote execution. \cref{sec:prob} presents an overview the system, formulates the problem and presents our solution. In \cref{sec:sim_result}, we present the experimental results. Finally, we conclude in \cref{sec:con}.

\section{Satellite Architecture} \label{sec:sat_arch}
In this section, we present a brief overview of the satellite architecture, the communication requirements, and present the current and future applications for satellite networks.
\subsection{Constellations} \label{subsec:const}
Traditionally, satellites were designed to operate in a solitary environment and their data flow followed a bent-pipe architecture where an earth station transmits the data to the satellite in the uplink. The satellite amplifies the signal and transmits it to another Earth station in the downlink. An example is the usage of geo-stationary satellite for voice calls
\cite{radhakrishnanSurveyInterSatelliteCommunication2016}. Due to their high altitude (36,000 km), the area covered by a geo-stationary satellite can be large. Hence, only a few of them are necessary to cover the entire earth. However, as their distance is large, the communication delay is large as well. Typical \gls{rtt} for a geo-stationary satellite can be more than $600$ ms \cite{quLEOSatelliteConstellation2017}. To minimise this delay, the satellites have to orbit the earth at a much lower altitude. However, this means only a fraction of the earth's region can be covered at any time and many satellites would be necessary for global coverage. For example, Iridium network system operating at $760$ km (\gls{leo}) requires $66$ satellites to cover the entire earth's surface \cite{nishiyamaLoadBalancingQoS2011}. Such a satellite formation can work solitary or in a constellation. According to \cite{radhakrishnanSurveyInterSatelliteCommunication2016,davoliSmallSatellitesCubeSats2019}, there are three common types of formations of satellites namely: 
\begin{itemize}
    \item Trailing: In this formation, satellites share the same orbit, but are separated by a specific distance.
    \item Cluster/ Swarm: Satellites in this formation fly in close proximity to each other but in their own orbits.
    \item Constellation: A set of satellites organised in different orbital planes that cover the entire earth. Reference \cite{kakLargeScaleConstellationDesign2019} presents a large scale constellation design framework for \gls{iost}.
\end{itemize}

\subsection{Communication}\label{subsec:communication}

{ The wireless communication refers to the transmission and reception of the data between a satellite and other entities. A satellite may communicate with } 
\begin{enumerate}
	\item other satellites and space station,
	\item earth bound entities, and
	\item planetary rovers.
\end{enumerate}
{The satellites may not only communicate to other satellites in the formations described above but also between the satellites in \gls{leo}, \gls{meo} and \gls{geo} orbits \cite{mukherjeeCommunicationTechnologiesArchitectures2013,akyildizMLSRNovelRouting2002}. The reader may refer to \cite{davoliSmallSatellitesCubeSats2019,mukherjeeCommunicationTechnologiesArchitectures2013,saeedCubeSatCommunicationsRecent2020,kodheliSatelliteCommunicationsNew2021} for a comprehensive surveys on inter satellite communication system.}
\subsection{Computing in space}\label{subsec:computing}
The small and nano satellite constellations can be used for various applications. For example, the satellite mesh can be used as a backhaul network providing high-speed, low latency communication links \cite{akyildizInternetSpaceThings2019}. It could be particularly useful in remote areas where the terrestrial network does provide coverage. Reference \cite{nishiyamaLoadBalancingQoS2011} shows how the network load can be anticipated based on the geographical location and uses communication links between \gls{leo} satellites, as well as \gls{leo} and \gls{geo} satellites to improve the \gls{qos} for the end users in their communication needs.

The small and nano-satellites can provide a computational platform in space \cite{songEnergyEfficientMultiAccess2021a,quLEOSatelliteConstellation2017a}. It could process data generated by itself. For example, satellites and their networks can be used for sensing the earth. \cite{akyildizInternetSpaceThings2019} referred to monitoring and reconnaissance capabilities of satellite networks as \textit{``Eyes in the sky"}. Satellites equipped with sensors such as cameras can monitor the earth's surface $24\times 7$. However, not all data that is acquired is useful due to various reasons. For example, if the interest is monitoring assets in an urban area, images of rural areas or oceans are of little interest. Also, if the images acquired are occluded by the clouds it could render them useless \cite{giuffridaCloudScoutDeepNeural2020}. Transmitting all the acquired images to the earth station can put a strain on the communication links as well as on the earth station. The problem would only exacerbate in the future when more satellites are launched. If the satellites could pre-process the images and only retain and transmit the useful images, the resources could be more manageable.  \cite{giuffridaCloudScoutDeepNeural2020} used various deep learning algorithms to filter out the images that are occluded by the clouds. 

In a different scenario, the authors in \cite{yuECSAGINsEdgeComputingenhanced2021} describe using satellites to provide continuous \gls{iov} services. In this case, the data generated by the vehicles on the surface of the earth can use edge-computing and communication services provided by the satellites to communicate with other vehicles in the \gls{iov}. The satellites not only act as a low latency communication medium but also a computing platform.
In this work, we assume the satellites to be resource-limited computation platforms and the proposed algorithms would benefit such systems.

\subsection{Security risks in Space}\label{sec:security}
In space, attacks can be physical/kinetic or cyber \cite{toddharrisonSpaceThreatAssessment2020}, and the impact caused by such attacks depends on the sophistication of the attackers. If the attacker is an individual with limited capabilities like a hacktivist or an insider with limited capabilities, the impact may be limited. If the threat actor is a hostile nation-state or a privileged insider, the impact will be significantly higher \cite{bradburyIdentifyingAttackSurfaces2020}.{ Consider a defunct satellite that is out of service but still in orbit. If a hacker gains access to such a satellite, they may launch an attack on other satellites and services.}
The major reasons for concerns are the following:
\begin{itemize}
\item \gls{cots} hardware and software may have reported flaws and threats. Satellites may be in orbit for a long period of time (years) by which time new vulnerabilities could be discovered. It may be impossible or financially infeasible to apply the patch or update the software in the orbit,
\item \glspl{sme} may overlook security in favour of cost-saving,
\item hackers and activists also have access to the same technology (hardware and software) as it is readily available.
\end{itemize}

\begin{figure*}
	\centering
	\includegraphics[width=.75\textwidth]{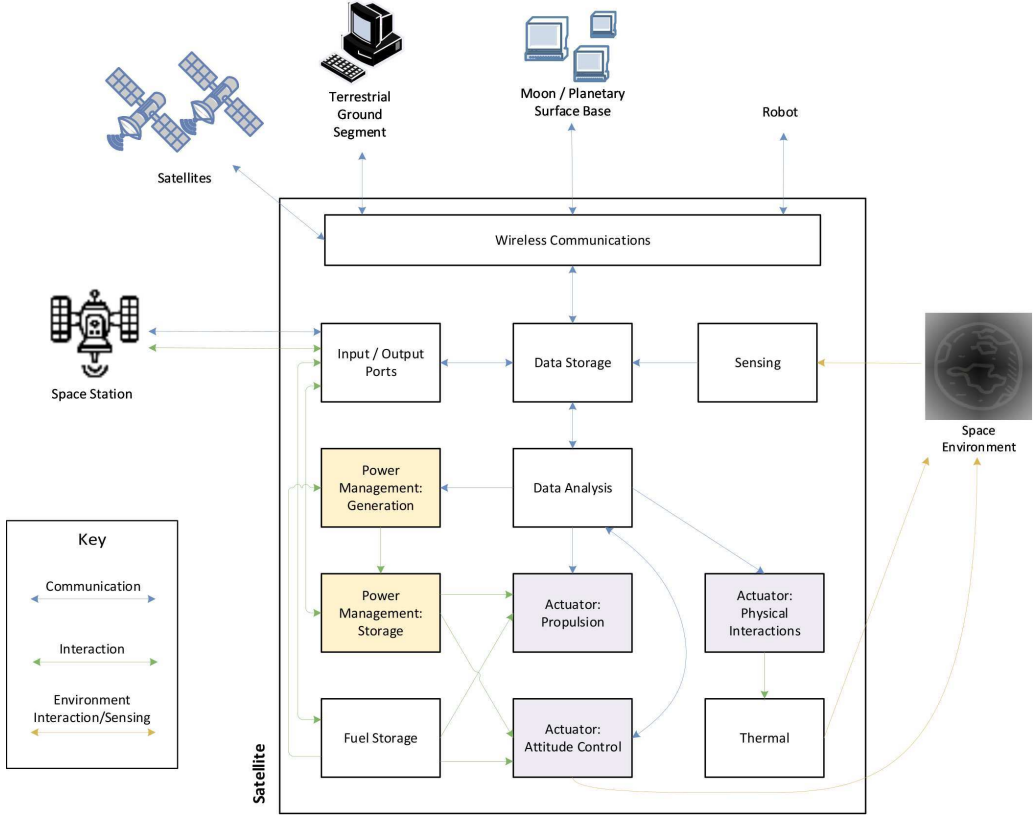}
 	\centering
 	\caption{Functional reference architecture of a satellite detailing its attack surface  \protect\cite{bradburyIdentifyingAttackSurfaces2020}} \label{fig:ref_arch}
\end{figure*}

{\gls{ra} is often used to understand and mitigate the security risks. They can be used in conjunction with attack trees for security-minded verification \cite{mapleSecurityMindedVerificationSpace2020}. \cref{fig:ref_arch} shows a functional \gls{ra} of a satellite operating in an orbit. It shows the functional blocks within the satellite and interfaces for it to interact with the external world. It also highlights the attack surface of the satellite such as the Input/Output ports that may be targeted in an attack. In this work, however, we focus our study on the attacks that may be directed toward wireless communication.
In general, cyber attacks affect one or more of the three aspects of security collectively known
as the \gls{cia} triad.}
\subsubsection{Confidentiality}
Data confidentiality refers to the protection of transmitted data from passive attacks such as eavesdropping \cite{10.5555/2523199}. {If confidential information is being shared without encryption or poor encryption, a passive attacker may listen to the communication or use data sniffing techniques to learn the victim's secrets \cite{alabaInternetThingsSecurity2017}. \cref{tbl:conf} details various encryption algorithms ranked such that the strongest and slowest algorithm has the confidentiality score of one and other encryption algorithms are relative to it \cite{xieSchedulingSecuritycriticalRealtime2006b,chenSchedulingWorkflowsSecuritySensitive2017,huangDeepReinforcementLearning2019}.
We assume the confidentiality score ($\gls{conf_level}$) to be directly proportional to the process rate (i.e. stronger encryption algorithm has higher security overhead on the processor). In \cref{sec:sim_result} we base our decision to select the appropriate encryption algorithm based on this table.}

\begin{table}
	\centering
	\caption{Encryption algorithms in literature, their security level and process rate \protect\cite{huangSecurityCostAwareComputation2019}}\label{tbl:conf}
	\RaggedRight
	\begin{tabular}{lrr}
		\textbf{Encryption}&\textbf{Confidentiality (\gls{conf_level})}&\textbf{Process Rate (Mb/s)}\\
		\hline
		IDEA& 1.0& 11.76\\
		DES& 0.85& 13.83\\
		Blowfish& 0.56& 20.87\\
		AES& 0.53& 22.03\\
		RC4& 0.32& 37.17\\
		
		\hline
	\end{tabular}
\end{table}
\subsubsection{Integrity}

The integrity of the data is compromised when the attacker modifies the data from the sender to the receiver.
{Attacks such as the man-in-the-middle attack can modify the data, and the receiver satellite may have no knowledge about it. In an extreme scenario, if the receiving satellite has a propulsion system, the attacker may modify control messages to move away from their orbit, burn its fuel unnecessarily, or fatally crash with other nearby satellites \cite{bradburyIdentifyingAttackSurfaces2020}. \cref{tbl:int} details a number of hashing algorithms to ensure the data has not been falsified. We used these values in \cref{sec:sim_result} to select the appropriate hashing algorithm.}

\begin{table}
	\centering
	\caption{Different hashing algorithms, their security level and process rate \protect\cite{huangSecurityCostAwareComputation2019}}\label{tbl:int}
	\RaggedRight
	\begin{tabular}{lrr}
	
		\textbf{Encryption}&\textbf{Integrity (\gls{int_level})}&\textbf{Process Rate (Mb/s)}\\
		\hline
		TIGER& 1.0& 75.76\\
		RipeMD160& 0.75& 101.01\\
		SHA-1& 0.69& 109.89\\
		RipeMD128& 0.63& 119.05\\
		MD5& 0.44& 172.41\\
		\hline
	\end{tabular}
\end{table}
\subsubsection{Availability}

Attackers and hackers may try to disrupt the service provided (by the servers) by employing \gls{dos} attacks or jamming the communication channel. {Similar to attacks on the Earth's surface, attackers can affect the availability of wireless channels by transmitting at the same time as the legitimate satellite. Satellites sharing the same channel would be affected. Similarly, the attacker may target the server by making too many requests so that the legitimate node cannot be served. We simulate such attacks on availability in \cref{sec:sim_result} by degrading the channel (--see \cref{fig:wifi}).}

\section{Computation Offloading}\label{sec:co}
\begin{table}
	\caption{List of important symbols.}
\begin{tabular}{lp{0.85\columnwidth}}
	\gls{job_data} &Job data size.\\
	\gls{Em}&Total energy usage for executing the job.\\
\gls{Eoff}&Total energy usage for offloading the job.\\
	\gls{Im}&Number of instructions a device can execute per second.\\
\gls{Is}&Number of instructions the server can execute per second.\\
	\gls{J}&Maximum number of jobs in an interval.\\
\gls{queue_servers}&Number of servers in queuing model equals number of
	cores in the satellite.\\
\gls{queue_servers_server}&Number of servers in queuing model equals number of
	cores in the space station.\\
	\gls{Pc}&Instantaneous power of satellite while executing the job.\\
\gls{comm_rate}&Data Rate.\\
	\gls{SD}&Desired security level.\\
\gls{SP}&Selected security level.\\
	\gls{job_compute_per_bit}&Number of instructions to execute a job per bit.\\
\gls{job_arrival_rate}&Job generation rate.\\
	\gls{action}&set of actions for the Markov Decision Process (MDP).\\
\gls{reward}&Reward function for the MDP.\\
	\gls{state}&set of states for the MDP.\\
\gls{transition}&Transition function for the MDP.\\
	\gls{t_deadline}&Maximum allowed time for completion of a job.\\
\gls{tm}&Time taken by satellite to execute the job.\\
\gls{risk}&Risk.\\
\gls{util}&Central Processing Unit (CPU) utilisation.\\
\gls{serv_rate}&Service rate of the node.\\
\end{tabular}   
\end{table}

To understand and mitigate the attacks on communication systems of satellites and space systems, we study a process called \acrfull{co}. In \cref{subsec:computing}, the application of satellites as a computing platform was described. However, in \gls{ns} systems, satellites may be constrained in terms of their computation and energy resource, \gls{co} will be very useful.
\gls{co} is a process of delegating a computationally intensive task to an alternative device rather than on its own computing platform. This delegation may be done for achieving various goals such as improving latency, conserving energy, or both. Many \gls{co} algorithms have been proposed to offload algorithms from the edge devices to the cloud known as \gls{mcc} as well as to the edge servers known as \gls{mec} \cite{shiEdgeComputingVision2016,maoSurveyMobileEdge24,abbasMobileEdgeComputing2018,
machMobileEdgeComputing23,sthapitDistributedComputationalLoad2017,minLearningBasedComputationOffloading2019}. In the context of \gls{ns}, a satellite may be considered as a resource-limited device that offloads some of its computation to the neighbouring satellites, space station or ground station to save resources. On the other hand, in the future, satellites may offer such computing services to \gls{ue} such as smartphones and vehicles, similar to services currently offered by cloud and edge servers. Especially with \gls{leo} satellites constellations such as OneWeb\footnote{\url{www.oneweb.world}}. and Starlink\footnote{ \url{www.starlink.com}} providing high-speed low-latency internet connectivity to worldwide coverage including remote areas not covered by cellular services. 

Consider a resource-limited satellite such as Cubesats equipped with a sensor that is operating in the orbit. The sensor senses its surroundings and readings are sent to the processor periodically. The readings have to be processed by a computationally intensive algorithm which requires computing and energy resources. The satellite can, however, offload the computing to a nearby satellite or space station which has a significantly higher computational capability as well as energy resources. Also, for simplicity consider both satellite and the space station are static in position relative to each other for the duration of offloading. However, the offloading is not always fruitful and depends on the connection quality to the space station. If the wireless channel is used by other satellites in the vicinity, transmitting sensor data can be a lengthy process. Similarly, if the space station is already busy with other algorithmic jobs, it may take a long time to service the satellite. Both of these delays may mean that the satellite misses the threshold time for completion of the algorithm. Missing the threshold is not desirable for the satellite and should be avoided as much as possible. Additionally, the communication between the satellite and the space station may not be secure. If there is a rogue (compromised) satellite in the vicinity trying to attack either passively or actively, they may launch attacks described in the \cref{sec:security}.

\subsection{Job}
A job is a computationally intensive algorithm that the satellite is trying to offload. For example, a pose estimation algorithm using camera images can be an offloading job. We define it as a tuple $<\gls{job_compute_per_bit},\gls{job_data},\gls{t_deadline}>$ where $\gls{job_compute_per_bit}$ is the number of computation cycles per bit required to complete the job, $D$ is the data requirement of the job, $\gls{t_deadline}$ is the latest time to complete the job \cite{maoSurveyMobileEdge24}. Such a job may have a large value for $\gls{job_compute_per_bit}$ and need significant time and energy resources to complete. In literature, jobs are considered to be offloadable or not offloadable as well as full or partial offloading \cite{maoSurveyMobileEdge24}. However, for simplicity, we consider all the jobs to be offloadable and only full offloading is considered. The jobs are generated on a regular basis as a Poisson process with a mean arrival rate \gls{job_arrival_rate}.

\subsection{Local Computation}
A job can be processed locally using the satellite's own computing platform. In terms of security risk, as it does not involve any communication. Thus, we assume that such local computations are risk-free. However, if the device is already busy, each job has to wait for its turn. We model the local computation using Queuing Theory. We consider jobs to be processed on a \gls{fcfs} without pre-emptive scheduling. The service times are dependent on the job itself. Also, when the job is offloaded with a certain security level (--see \cref{tbl:conf,tbl:int}) it will impact the processing resources depending on the security level selected. As the service times for job and security levels can be significantly different, they can be modelled as hyper-exponential distribution \cite{stewartProbabilityMarkovChains2009}. Lastly, the number of queue servers (\gls{queue_servers}) is the number of cores in the device.

\subsubsection{Time}
Time taken by a satellite to execute a job is given by
\begin{equation}
\gls{tm}=\frac{\gls{job_compute_per_bit}\times \gls{job_data}}{\gls{Im}} \label{eqn_time}
\end{equation}
where \gls{Im} is the capability of the satellite to execute instructions usually measured in \gls{mips}. Before execution, there is a waiting time due to queuing which can be estimated using Little's law. Although, \gls{Im} may change depending on several factors including \gls{DVFS} we consider a fixed policy such that the \gls{Im} does not change over the time.
\subsubsection{Energy} 
The \gls{CPU} power is made up of two parts, the idle power and the running power, as
follows:
\begin{equation}
	\gls{Pc}=\gls{util}*P_{max}+(1-\gls{util})*P_{idle}
\end{equation}
where, $\gls{util}$, $P_{max}$, $P_{idle}$ are the utilisation, maximum power and idle power consumption of the \gls{CPU} respectively. So, energy consumed to execute a job can be calculated as 
\begin{equation}
\gls{Em}=\gls{Pc}\times\gls{tm}. \label{ch4:eqn:e_onboard}
\end{equation}

\subsection{Remote Execution}
The remote platform could be the other satellites, space station or ground station. We assume the server has \gls{queue_servers_server} cores available for computing, each core capable of executing \gls{Is} \gls{mips}. Before an algorithm can be executed on the remote platform, however, the data (possibly code as well) has to be transferred to the remote platform. However, there is a risk that one of the \gls{cia} aspects is breached while communicating. So, appropriate levels of security have to be put into place. Depending on which encryption algorithm and hashing algorithm is selected (--see \cref{tbl:conf,tbl:int}), different levels of security can be maintained. Also, different algorithmic complexity of the algorithm means they will incur different times and energy costs which are described below. Similar to other algorithms in the literature, we ignore the cost of sending the result back to the device as the data is of relatively lower size in most cases.  
\subsubsection{Time}
\begin{figure}
\centering
\includegraphics[width=0.6\columnwidth]{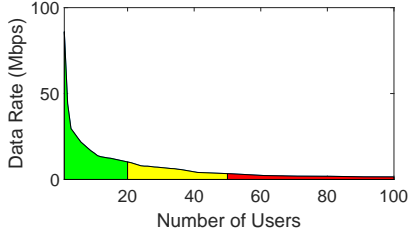}
\caption{Relationship between number of users and the data rate. Green region shows best network condition, yellow shows moderate and red signifies poor network condition. Availability is accounted in the model by considering poor network condition when the channel is jammed.    } \label{fig:wifi}
\end{figure}

The total time for executing a job on a remote platform can be estimated as follows:
\begin{equation}
\tau_{s}=\tau_{security}+\tau_{comm}+\tau_w  \label{eqn_total_time}
\end{equation}
where $\tau_{security}, \tau_{comm}, \tau_w $ represent the times taken by the \gls{off} to secure, packet, send the data, and wait for receiving the result respectively. {In space, the server may not be as powerful as the cloud on Earth. Hence there may be queuing as well. We model the delay in queuing with a queue similar to the queue in satellite. This is represented by $\tau_w$.} 
Similarly, if $\gls{job_data}$ is the data size to be transferred/received, the communication time is given by:
\begin{equation}
\tau_{comm}=\frac{D}{\gls{comm_rate}}
\end{equation}
where \gls{comm_rate} is the available data rate. However, \gls{comm_rate} may be shared between other users and effective bandwidth depends on other users' action. For a device $k$ within $N$ users sharing the communication channel, effective data rate can be calculated as
\begin{equation}
\gls{comm_rate}_{k}=Bw \log_2\left( {1+\frac{G_{k,k}.P_k}{\sigma^2+\sum_{i=1,i\ne k}^{N}G_{i,k}.P_i}}\right) \label{eqn_comm}
\end{equation}
where, $Bw$ is the bandwidth, $G_i,P_i$ are the channel gain and transmit power for user $i$. As evident from \cref{eqn_comm}, communication data rate depends on the channel gain and transmission power, as well as other users transmitting simultaneously. Herein, `other users' may include any devices trying to communicate on the channel. However, they could also be \textit{jammers} trying to disrupt the communication between a user and the base station. In that case, the availability is affected. We account for the availability by considering the jammed channel as poor network condition. Instead of calculating the communication rates using the transmit power and the channel gain, we use a simpler version based on arbitrary values in \cite{doi:10.1002/ett.3493}. It only depends on the number of users actively using the channel and drops off exponentially as the number of users increases. The rates for different numbers of users are visualised in \cref{fig:wifi}.{ The actual number of satellites in the vicinity in space is random at any point in time. However, if we consider the satellites are uniformly distributed around the globe, choosing a smaller satellite formation means a lower number of satellites actively communicating on average. Similarly, if we consider a larger satellite constellation, the average number of satellites actively involved in communication may be higher. }
\subsubsection{Energy}
The energy consumed for offloading can be calculated as:
\begin{equation} 
\begin{split}
\gls{Eoff}&=P_c \times \tau_{comm} +P_{max} \times \tau_{security}.\\
\end{split}
\end{equation}
\subsubsection{Security Risk}
\begin{figure}
    \centering
    \includegraphics[width=0.6\columnwidth]{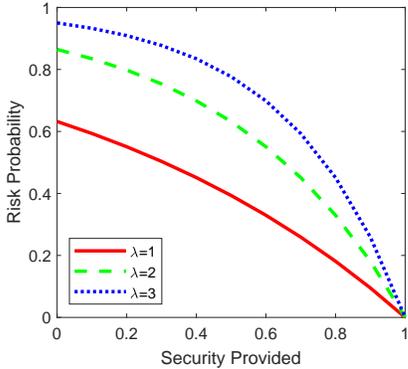}
    \caption{Probability of security breach for \gls{SD}=1}
    \label{fig:risk_prob}
\end{figure}
The security risk increases when \gls{co} is implemented because the system is relatively more vulnerable than if the computation is done locally. Some risks can be mitigated by choosing appropriate security measures such as encryption. Similar to \cite{xiaoyongNovelSecurityDrivenScheduling2011a,huangSecurityCostAwareComputation2019} we model the risk as Poisson distribution. 
\begin{align}
	P_k= \begin{cases}
		0,& if~ \gls{SD} \leq \gls{SP}\\
		1-\exp^{-\lambda^k(\gls{SD}-\gls{SP})}& if~\gls{SD}>\gls{SP}
	\end{cases}
\end{align}
where $k \in \{C, I\}$ is the particular security concern, $\gls{SD}$ is the security demand of the job, and $\gls{SP}$ is the chosen security level. If the security provided is greater than or equal to the security demand, then the risk is zero. However, if the chosen security is less than the required level, it is prone to security breaches. The exact probability of risk depends on $\lambda_k$ which can be different for each server as well as for confidentiality and integrity. The total probability of risk such that either confidentiality or integrity is violated is then given by
\begin{equation}
	P_r=1- \prod_{k\in\{C,I\}}(1-P_k).    \label{eqn:risk}
\end{equation}

\section{Problem Formulation}\label{sec:prob}
{ Recall that the state of the system in our problem corresponds to the number of jobs waiting in the queues at the local queue (at the satellites) and the number of jobs waiting at the server. For both these queues, the number of jobs in the next time slot will only depend on the number of jobs awaiting during the current time slot and the decisions (offloading/ local computation) taken during the current time slot. Note that for the local queues at the satellites, the number of jobs departing the task queue will depend on the decision taken during the current slot, whereas the number of new jobs arriving is assumed to be an independent and identical Poisson process (memoryless). Similarly, for the queue at the server, the number of new jobs arriving will depend on the decision taken during the current slot, whereas the departure process only depends on the computational time at the server. Note that splitting a Poisson process randomly with a fixed probability creates two separate Poisson processes. Similarly, if two Poisson processes are combined (for example two satellites may offload to the server in the same time slot) it results in a Poisson process \cite{stewartProbabilityMarkovChains2009}.
 Based on these observations, we note that, given the current state and action, the next state of the decision process is conditionally independent of all previous states and actions; in other words, the state transitions satisfy the Markov property. } A \gls{mdp} is a tuple $<\gls{state},\gls{action},\gls{transition},\gls{reward}>$  where $\gls{state}$ is a finite set of states, $\gls{action}$ a finite set of actions, $\gls{transition}$ a transition function defined as $\gls{transition}:\gls{state}\times \gls{action} \times \gls{state} \rightarrow [0,1]$ and
$\gls{reward}$ a reward function defined as $\gls{reward}:\gls{state}\times \gls{transition}\times
\gls{state} \rightarrow \mathbb{R}$ \cite{vanOtterlo2012}. {We also note that our setup is similar to existing works on computational offloading \cite{truong-huuOffloadWaitOpportunistic2014,zhangOffloadingMobileCloudlet2015,abualsheikhMarkovDecisionProcesses23}, which also model the problem as a \gls{mdp} and solve it using \gls{rl} methods.} An \gls{rl} agent observes the states at discrete intervals and makes the decision for the next time interval. 

{Recently, there has been a growing interest in applying \gls{rl} to the data and \gls{co} problem in terrestrial mobile networks. For instance, the problem of minimising the mobile user's cost, energy consumption and computation delay by offloading tasks to a mobile-edge computing server was considered in \cite{LiOffload2018} and \cite{RLOffloading2019}, and solved using Deep \gls{rl} techniques. \gls{rl} has been used to solve \gls{co} problem in \gls{iot} devices with energy harvesting as well \cite{minLearningBasedComputationOffloading2019,maoDynamicComputationOffloading2016}. The problem of allocating computing and network resources under varying \gls{mec} conditions was considered in \cite{WangMEC2019}. 
Reference \cite{GuVNF2020} applied DRL to solve the network utility maximisation problem in a \gls{vnf} environment. 
For a detailed survey on the application of \gls{rl} in \gls{co} in wireless networks, we refer the reader to \cite{DRLSurvey2019}. However, none of these works 
focus on \gls{co} in a satellite environment and the corresponding domain-specific features.}


For this problem, we consider the number of jobs in the queue of the satellite, the number of jobs in the server, the number of satellites communicating in the current time slot, and the number of jobs arriving in the time slot as the observations of the system. The satellite will not always know the exact number of jobs the space station is serving. However, when the server may agree to serve the satellite, it may send regular updates on its state. In our previous work we used a proactive and reactive algorithm to send this information (\gls{NSI}) about one's state to neighbours \cite{sthapitComputationalLoadBalancing2019}. 
The number of satellites communicating in a given time slot provides an inclination to the available bandwidth similar to the feedback channel gain. For the simulation, we assume that the number of satellites using the channel is random and independent of the previous interval but constant throughout the interval. Let $P_{off}$ be the probability of offloading to the server. Then, the time consumption to execute a job can be estimated using \cref{eqn_time,eqn_total_time} as follows:
\begin{equation}
    \tau=P_{off}\tau_m + (1-P_{off})\tau_s. \label{eqn:total_t}
\end{equation}
Similarly, energy consumption can be estimated as
\begin{equation}
    E=P_{off}E_m + (1-P_{off})E_s. \label{eqn:total_e}
\end{equation}
While the time and energy consumption can be estimated from the system state such as the number of jobs in the \gls{CPU}, communication, and the server queues, security risk cannot be observed directly or in advance. However, given enough data on previous observations and cost, we can estimate the risk conditions if it is time-invariant. The overall cost of executing is then given by
\begin{equation}
\begin{split}
	C_t&= \sum_{j=1}^{\gls{J}}( w_t \tau_j +w_e E_j + w_r \gls{risk}_j) \label{eqn:total_cost}\\
	subject~to ~ \tau &\le \gls{t_deadline},\\	
	\end{split}
\end{equation}
where, \gls{J} is the maximum number of jobs in an interval, $\tau_j,~E_j,~ \text{and } \gls{risk}_j$ are time, energy and risk while executing job $j$. $\gls{risk}_j$ is the random value sampled using \cref{eqn:risk} to represent the risk. $w_t,w_e,w_r$ are the weights for time, energy and risk components. Next, we relax the hard constraint on the time deadline to a soft constraint such that if the constraint isn't met, we add a large cost to the cost function whereas when the constraint is met the weight is zero. Also, as we propose a generic solution, we do not set these weights to custom values. Instead, we set them to equal weights. For applications that are specific, the weights could be adjusted to the application. For example, in a satellite communication network, when there are not enough satellites, the destination may not reachable and the data packets may be dropped. For such applications \gls{dtn} routing protocols are used \cite{madniDTNNonDTNRouting2020}. When \gls{co} used on such protocol, the weights on time can be set to zero.
\begin{equation}
    C_t= \sum_{j=1}^{\gls{J}}( w_t \tau_j +w_e E_j + w_r R_j+w_{d}(\tau_d-\tau))\\ \label{eqn:newcost}
\end{equation}
where,
\begin{equation}
w_{d}=\begin{cases}
0, & if~ \tau_d>\tau     \\
\text{non negative number},& if~ \tau_d\le \tau.
\end{cases}
\end{equation}

Our objective is to minimise \cref{eqn:newcost} in the long term. 

\begin{equation}
   \argmin_{P_{off},SL_{C},SL_{I}}	\mathop{\mathbb{E}} \left[\sum_{t=0}^{\infty} \gamma^t C_t\right] \label{eqn_problem}
\end{equation}
where, $\gamma$ is the discount factor. Our action is a three dimensional vector $[P_{off},SL_{C},SL_{I}]$ consisting of the probability to offload a job and the security levels to select to maintain the confidentiality and integrity of the offloaded data. Similar to \cite{sthapitComputationalLoadBalancing2019}, we select the execution platform probabilistically. The benefit of making such a decision is that the system does not need to know the exact number of incoming jobs. 

\subsection{\acrfull{ddpg}}
We use \gls{ddpg} \cite{lillicrapContinuousControlDeep2019} to solve the optimisation problem stated in \cref{eqn_problem}. \gls{ddpg} is a actor-critic based offline method that uses two separate \glspl{DNN} to approximate the Q-value network. Its main advantage over \gls{dqn} is its ability to work on a continuous action space \cite{Lillicrap2016ContinuousCW}. So the probability to offload $P_{off}$ can be any value ranging from $0-1$ and need not be discretised.
We trained our reinforcement agent on episodes of simulated data with each episode lasting $40$ seconds. We used the experience replay method for batch training and ADAM optimiser for the training purpose \cite{kingmaAdamMethodStochastic2017}. We trained the network for a maximum of $1000$ episodes. To stop the agent from being greedy and making sub-optimal decisions, we use the $\epsilon$-greedy approach whereby the agent makes a random action with a small probability $\epsilon$ and the rest of the time take the best (or greedy) action.  To balance the exploration and exploitation for the agent, we gradually lowered the value of $\epsilon$.

\subsection{\acrfull{dqn}}
\gls{dqn} \cite{mnihHumanlevelControlDeep2015} is the first reinforcement learning algorithm to demonstrate human level performance on Atari games. \cite{huangSecurityCostAwareComputation2019} used \gls{dqn} based algorithm to create security aware \gls{co} algorithm. In order to compare the performance of our proposed \gls{ddpg} based algorithm, we implemented a similar \gls{dqn} based solution for our problem. As \gls{dqn} works with a discrete action space, we quantised $P_{off}$ with a resolution of 0.2 ranging from 0 to 1. Otherwise, the other training parameters were left same as the \gls{ddpg} algorithm described below.

\subsection{Static policies}
We compared our proposed \gls{ddpg} algorithm against three static policies defined below.
\subsubsection{\gls{lo}}As the name suggests, it cannot offload any job and is oblivious to network changes and risk states.
\subsubsection{\gls{sons}}
This policy offloads all the jobs to the server without following any security guidelines. So when the risk is high, attacks are always successful.
\subsubsection{\gls{soms}}
Similar to the previous policy, it offloads all the jobs. But, it uses the highest security measures regardless of the network conditions.

\section{Simulation Results}\label{sec:sim_result}
\glsresetall

In this section, we briefly explain our simulator, parameter selection and their results. We created our own simulator; all the code and environment are available at \url{https://github.com/sausthapit/ComputationOffloadingRL}. We used Matlab and Simulink environment which provides toolboxes for \gls{rl} and event-based simulations. The simulator also supports \gls{rl} agents with discrete action spaces such as \gls{dqn}.

{ We assume the maximum number of satellites will be different in the three formations due to the physical setup. In particular, the trailing formation occupies the least space (as the satellites share the same orbit) and has the least number of satellites. In comparison, the swarm formation has more satellites (as it involves satellites in different orbital planes). Lastly, the constellation formation has the largest number of satellites taking part in the communication (to cover the entire earth). According to  \cite{radhakrishnanSurveyInterSatelliteCommunication2016}, the transmit power for inter-satellite communication, $P_k,$ is in the range 0.5~W to 2~W. In the trailing formation, since the satellites are in the same orbit (and hence close to each other and less interference), we assume a lower transmit power and set $P_{comm} = 0.5$~W for the trailing formation. As the swarm and constellation formations occupy progressively larger areas of space, we assume $P_{comm} = 1$~W for the swarm formation, and $P_{comm} = 2$~W for the constellation formation. By default, we chose the swarm formation for the rest of the simulation unless specified and} we set the following parameters for the simulation. The idle power ($P_{idle}$), execution power ($P_{max}$) of the satellite is set to 0.1 and 5 watts respectively, processing capability of the satellite (\gls{Im}) is set to $2.5\times 10^9$ \gls{mips} with $\gls{queue_servers}=4$. Similarly, we set the processing capability of the server satellite to be twice that of the satellite. Also, the number of cores in the server is higher than the number of cores in the satellite (i.e. $\gls{queue_servers}<\gls{queue_servers_server}=16$). The job size is chosen to be $0.2$ MB, and it takes one second to process on the satellite without the waiting times. For stability, the queues are limited to finite buffers. The maximum queue length for the \gls{CPU} and communication buffer is set to $20$ whereas, for the server, the computation buffer is set to $10$. Also, \gls{t_deadline} is set to $5$ seconds. This means if a job has to wait more than $5$ seconds to process it is not useful and is considered a dropped job. Similarly, if any of the queue buffers are full when a new job arrives, it is lost as well. In default settings, we consider on average three jobs arriving per second and best network setting and lowest risk level. For the training purposes, we set the weights $w_t, w_e, w_r, w_d$ to $1,1,10$ and $10 $ respectively. This implies whilst we would like to improve on execution times and energy, we would like to avoid losing jobs and minimise the security attack. In fact, setting $w_r$ and $w_d$ to the same large value suggests that a successful attack is as bad as dropping a job. However, the weight selection is done arbitrarily and can be tailored to the application requirement. For applications that are super sensitive to security threats, it could be even higher. Also, when measuring the performance instead of evaluating the actual time elapsed for each job, we count how many jobs were completed within the time threshold.
Once the agent is trained, we ran $10$ Monte-Carlo simulations with random seeds for each of the settings described below.
\subsection{Incoming job rate} 
\begin{figure}
    \centering
    \begin{tabular}{cc}
    \includegraphics[width=\figurewidth\textwidth]{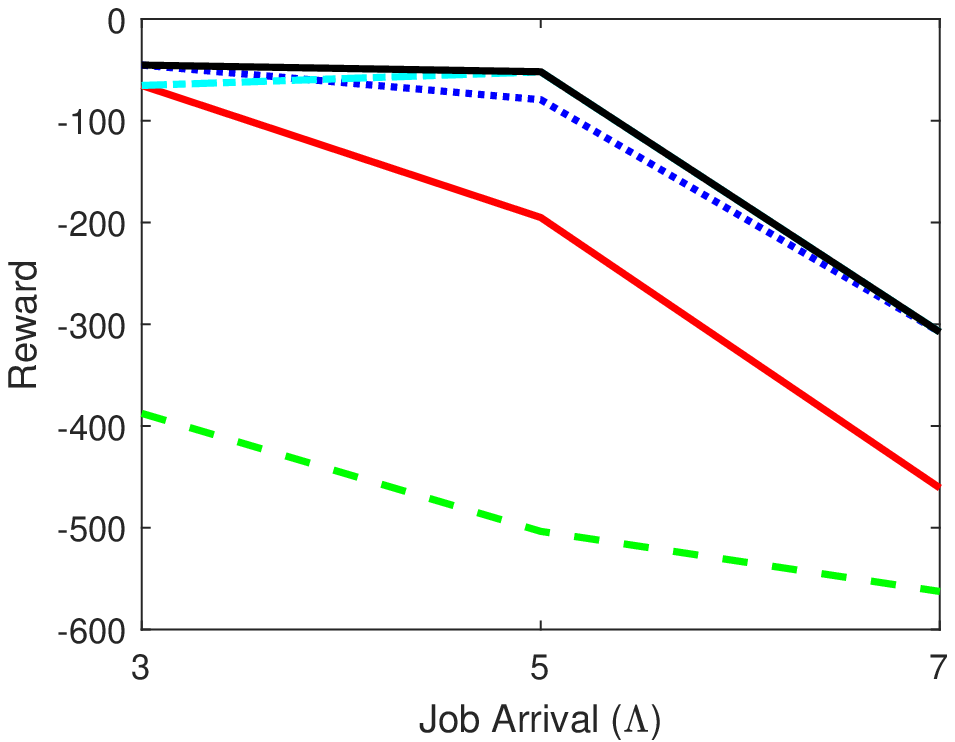}&
        \includegraphics[width=\figurewidth\textwidth]{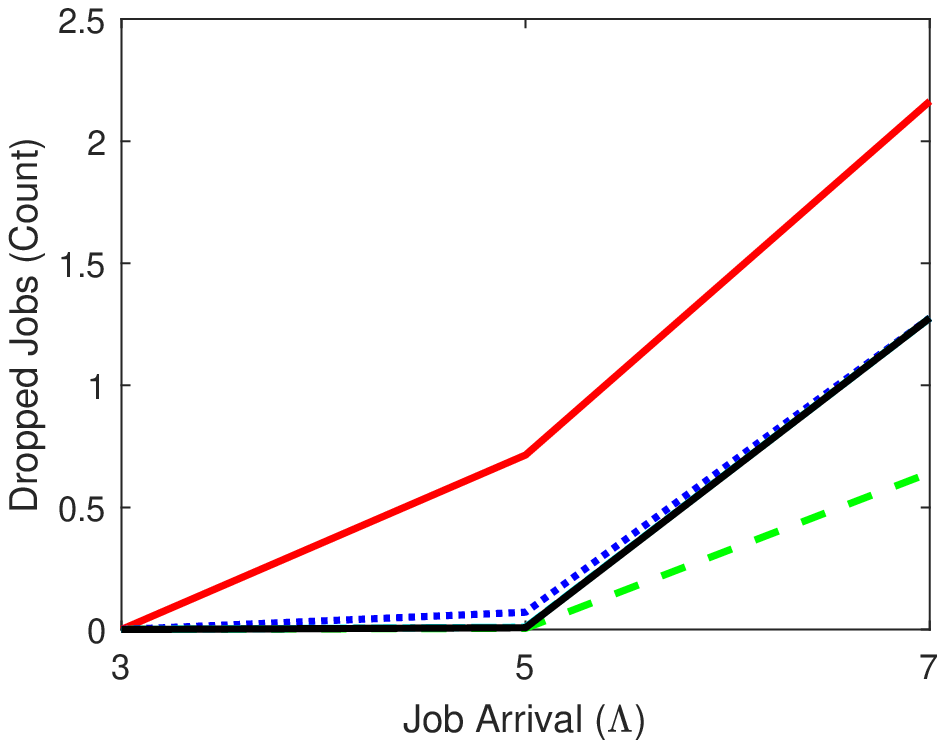}\\
        \includegraphics[width=\figurewidth\textwidth]{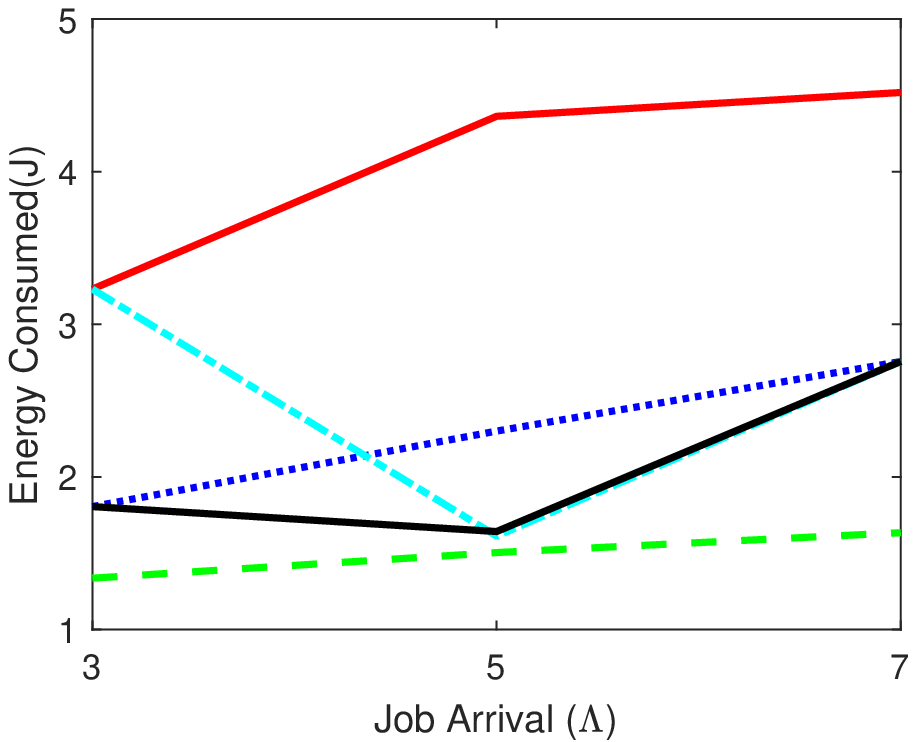}&\includegraphics[width=\figurewidth\textwidth]{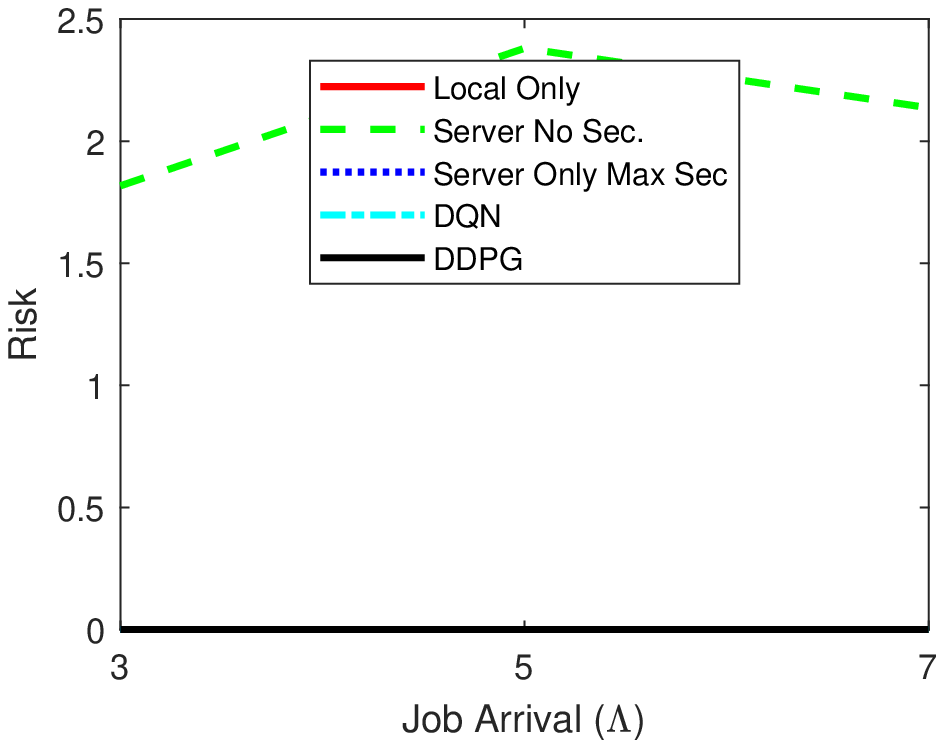}
    \end{tabular}
    \caption{Performance for different job arrival rates. Clockwise from top left:  average overall score, average dropped jobs, lost jobs due to security, and energy consumed.}
    \label{fig:arrival_rate}
\end{figure}

We simulated various job rates ranging from three jobs per second to seven jobs per second. If the satellite is equipped with a quad-core processor with a service rate of one job per second, when processing locally, its utilisation is given by 
\begin{equation}
    \gls{util}=\frac{\gls{job_arrival_rate}}{\gls{queue_servers}\gls{serv_rate}}=\frac{4}{4\times1}=1.
\end{equation}
Hence, it is only stable for $\gls{job_arrival_rate}<4$. Otherwise, the queue length will continue to grow indefinitely; in this case, as the buffer is limited, lost. The device is forced to offload or drop some of its jobs to maintain stability when the job rate is higher. The network condition is set to the best, risk level to the lowest, and data size to 0.2 MB. \cref{fig:arrival_rate} shows the averaged results for our experiments. The top left figure shows the overall cost achieved by each of the four policies. It is evident that the cost is growing for all the policies and \gls{sons} fared the worst. This is due to the security attacks it has suffered the most which are seen from the bottom right image. In terms of jobs dropped (-- see \cref{fig:arrival_rate}, top-right), all algorithms were able to process all the jobs at $\lambda=3$. However, as the arrival rate started to increase, the local computation suffered and dropped the most jobs averaging more than two jobs per episode. Both \gls{dqn} and \gls{ddpg} have similar results with \gls{dqn} using slightly more energy at lower job arrival setting. 

\subsection{Network Environment}
\begin{figure}
    \centering
    \begin{tabular}{cc}
    \includegraphics[width=\figurewidth\textwidth]{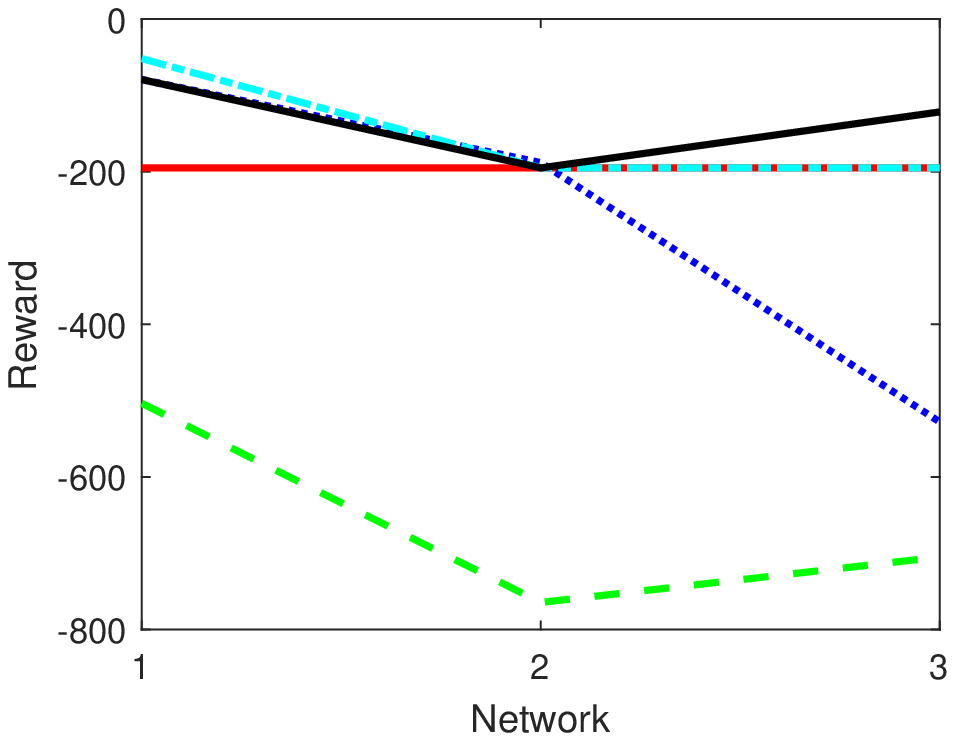}&
        \includegraphics[width=\figurewidth\textwidth]{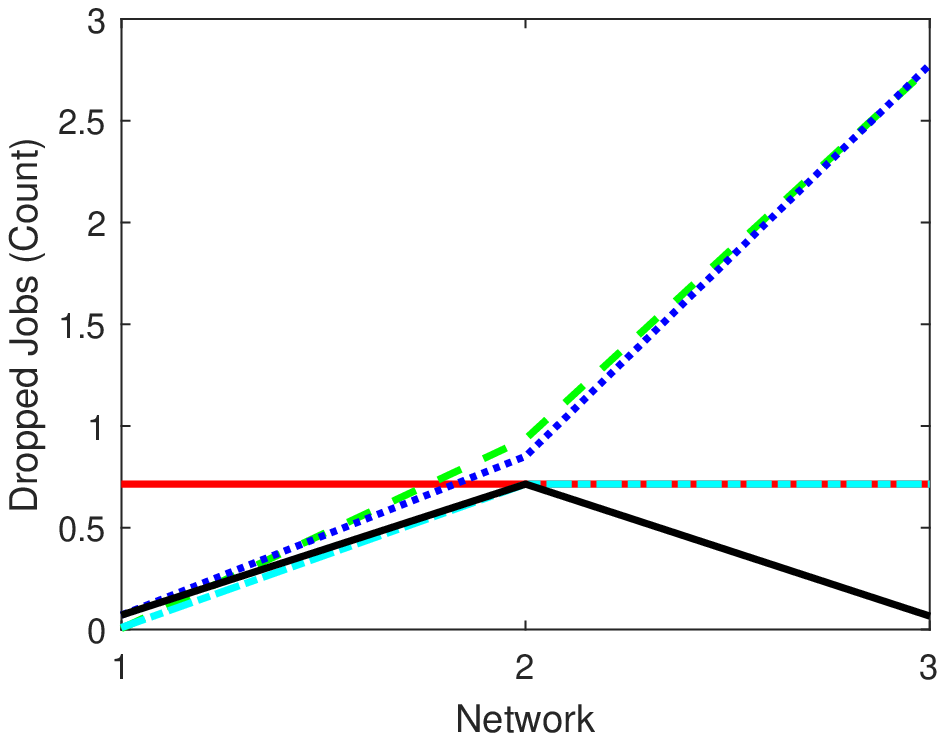}\\
        \includegraphics[width=\figurewidth\textwidth]{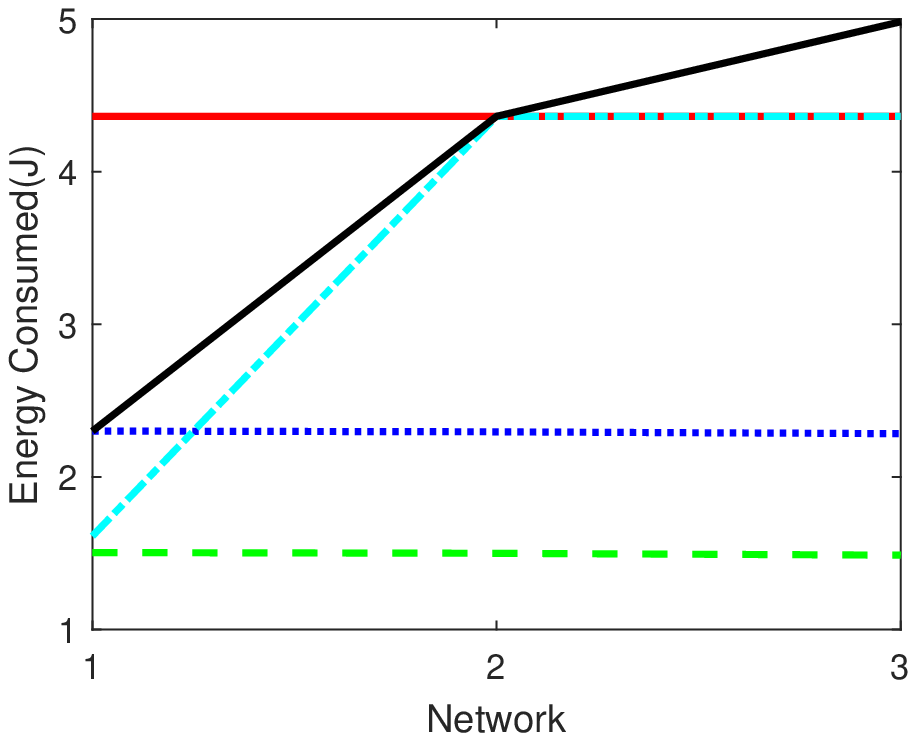}&\includegraphics[width=\figurewidth\textwidth]{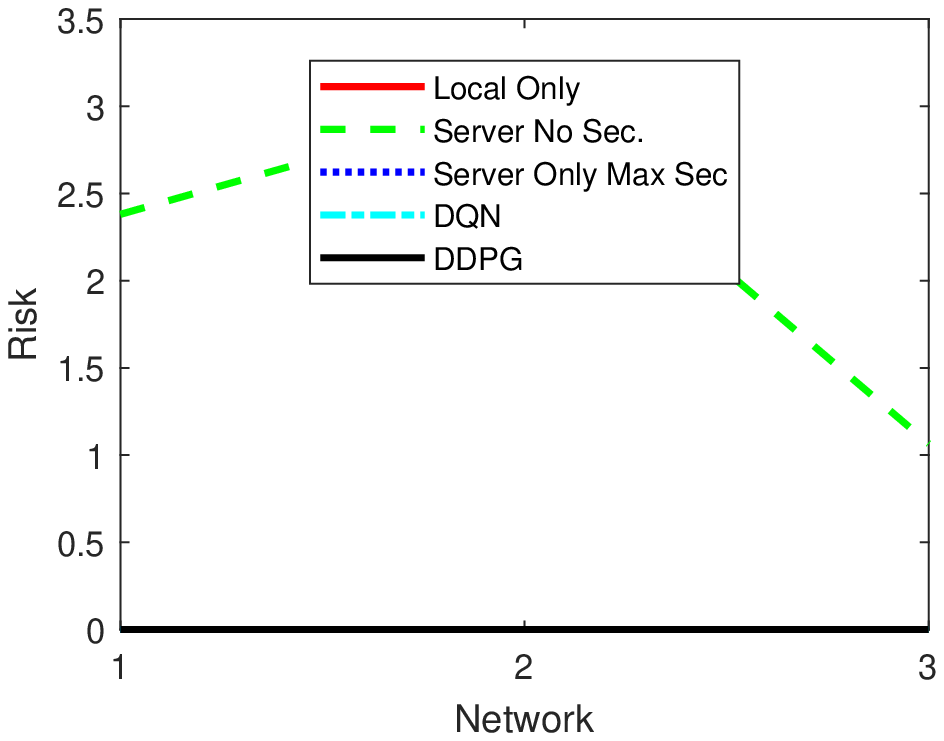}
    \end{tabular}
    \caption{Performance for different network conditions; $1,2,3$ represents best, medium and worse conditions. Clockwise from top left:  average overall score, average dropped jobs, lost jobs due to security, and energy consumed.}
    \label{fig:net_conditions}
\end{figure}

We considered three network settings, namely best, medium and poor represented by green, yellow and red area in the \cref{fig:wifi}. In the best setting (which is the default setting), only a few users simultaneously communicate at a given time slot, whereas in the medium setting, considerably more users communicate at the same time. Poor settings may represent a large number of satellites communicating at the same time or a malicious attacker trying to deliberately jam the channel. To simulate these settings, we simply use a uniform random number generator with boundary limits. Results for a varying network environment is presented in \cref{fig:net_conditions}. As expected the \gls{lo} policy is not affected by the varying network condition evident by the horizontal solid red line. Both \gls{sons} and \gls{soms} policies dropped similar amount of jobs per episode as seen in the \cref{fig:net_conditions} top right. This is because the satellite is unable to reach the server as the network condition worsens. However, the cyan dashed line for the \gls{ddpg} algorithm shows that even it performed better than the \gls{lo} algorithm suggesting that it used both local and remote resources in an efficient manner. This is evident from the bottom-left figure where the cyan line is using the most energy (up to $5$J at the worst network condition). In theory, the \gls{dqn} should also follow a similar pattern as \gls{ddpg} but in this case, when the network worsened, \gls{dqn} only used the local resources. 
\subsection{Risk Levels}
\begin{figure}
    \centering
    \begin{tabular}{cc}
    \includegraphics[width=\figurewidth\textwidth]{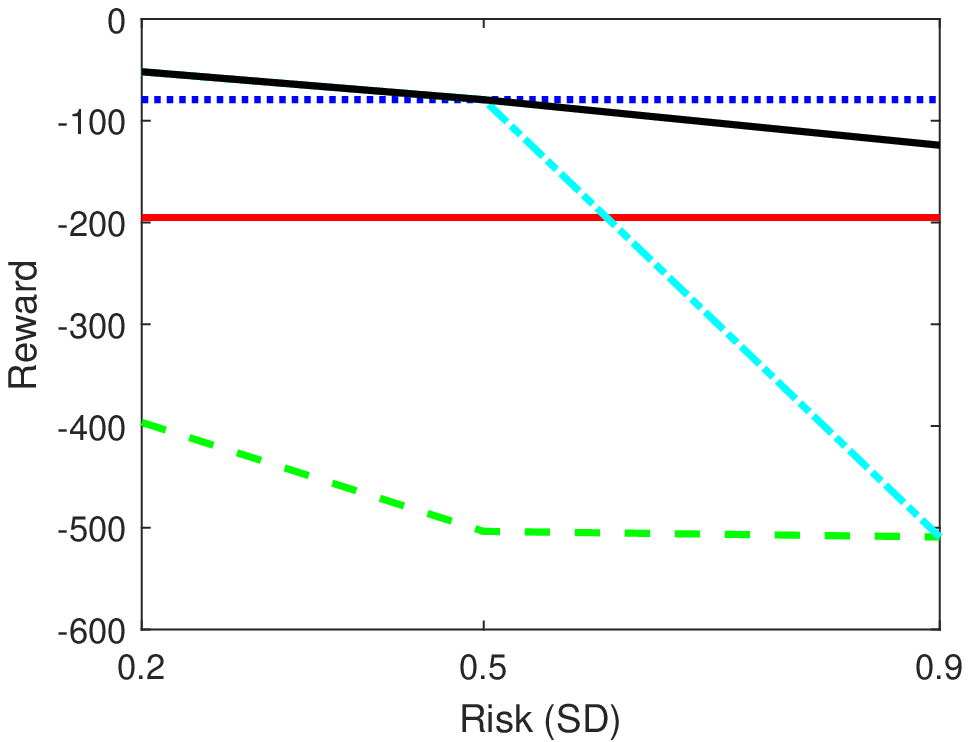}&
        \includegraphics[width=\figurewidth\textwidth]{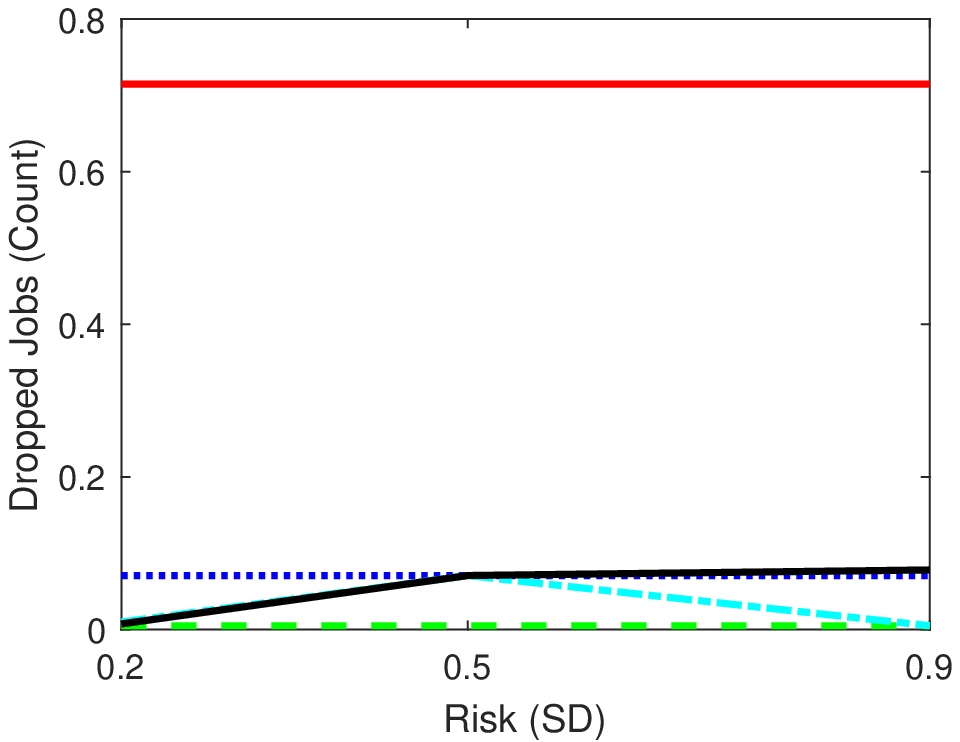}\\
        \includegraphics[width=\figurewidth\textwidth]{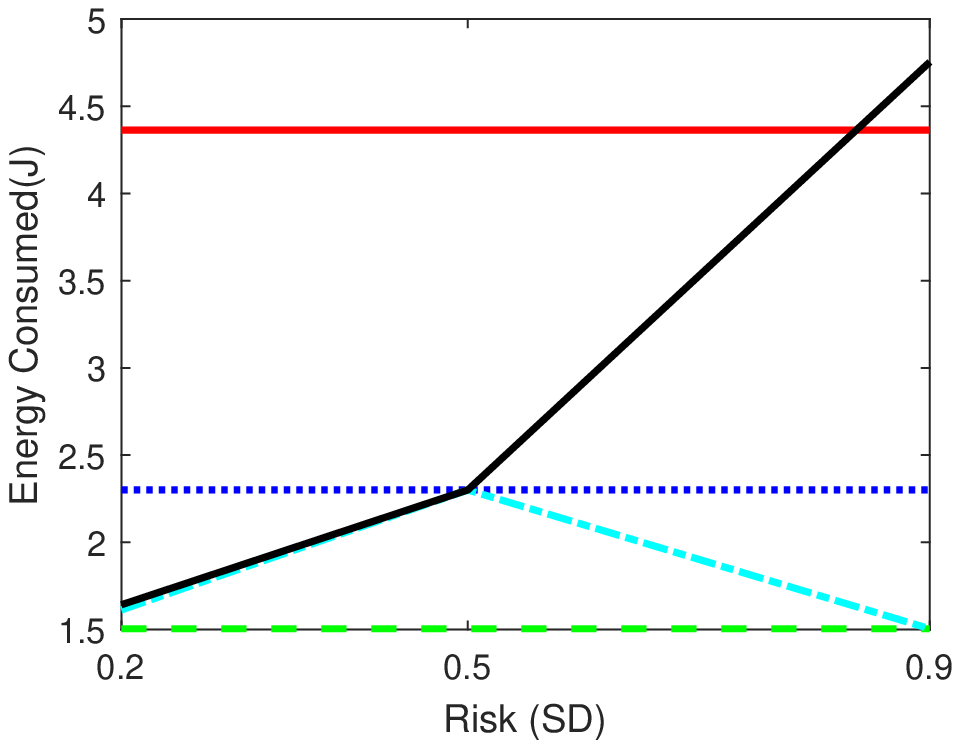}&\includegraphics[width=\figurewidth\textwidth]{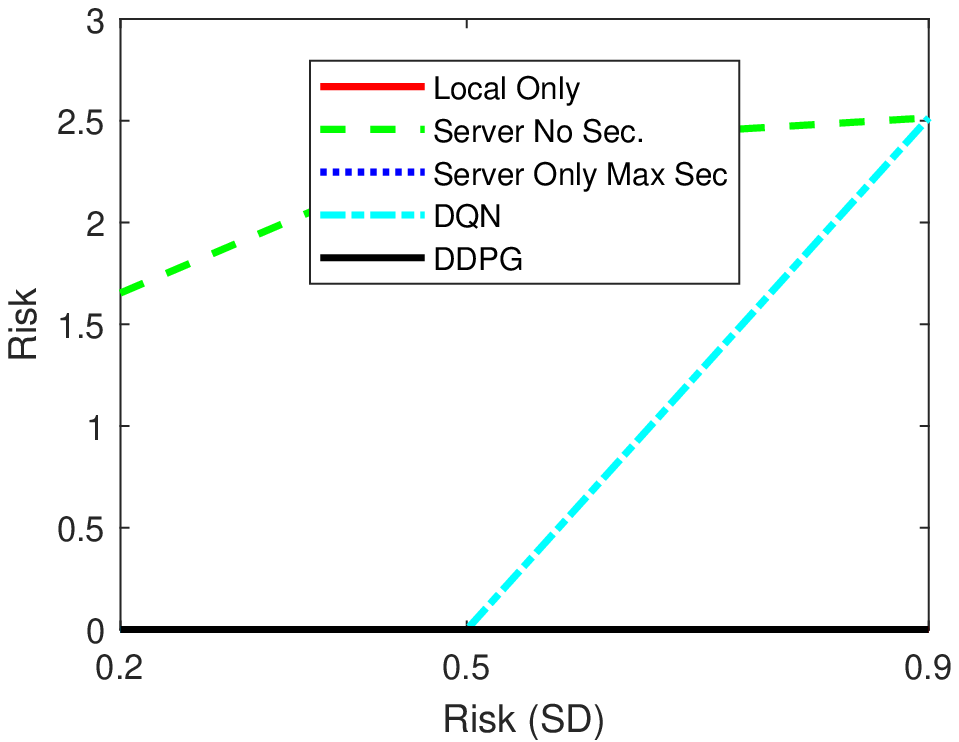}
    \end{tabular}
    \caption{Algorithm Performance for different risk conditions. Clockwise from top left:  average overall score, average dropped jobs, lost jobs due to security, and energy consumed.}
    \label{fig:risk_conditions}
\end{figure}

In this work, we modelled and simulated risk to emulate the real scenario of security attacks. In doing so, we changed the security desired ($SD$) parameter. The $SD$ sets the security threshold that needs to be fulfilled to save from attacks. While \gls{soms} is helpful to prevent unwanted security attacks, it uses vital computational resources, time and energy. While this may not be a problem when the resources are adequate for example on the Earth's surface with a substantial processor and mains-powered device. It can be significantly crucial to save energy and resource in space. Using remaining resources like batteries may mean the satellite or rover is completely out of service. In order to avoid this scenario, it is crucial to save as much energy as possible. Our \gls{ddpg} algorithm in this instance is able to adapt to the varying security level in the environment without directly sensing it and only based on the previous results. \cref{fig:risk_conditions} shows the performance of all four policies. As usual, \gls{sons} is the only one subject to successful attacks. The \gls{ddpg} algorithm used less energy than the \gls{soms} algorithm when the security threat is pretty low ($\le 0.5$) as seen in the bottom-left image. However, we also notice that when the threat is significantly high (0.9) the proposed \gls{ddpg} algorithm did not offload to the server and did most of the work itself using significantly higher energy than the \gls{soms} algorithm. \gls{dqn} algorithm on the other hand was able to handle more jobs even when risk was the worst (-- seen in \cref{fig:risk_conditions} top left) although more jobs were subject to security attacks. Cases such as these can be investigated further to reason why a particular agent is taking such action. One way of teaching the agent would be by changing the $w_e$. Furthermore, the weights could be adjusted or different agents could be combined at different environmental settings. For instance, including the remaining fuel or battery resources into the agent's observation. This way the agent can act intelligently and decide whether to prioritise security or energy resources.

\subsection{Data size}
\begin{figure}
    \centering
    \begin{tabular}{cc}
    \includegraphics[width=\figurewidth\textwidth]{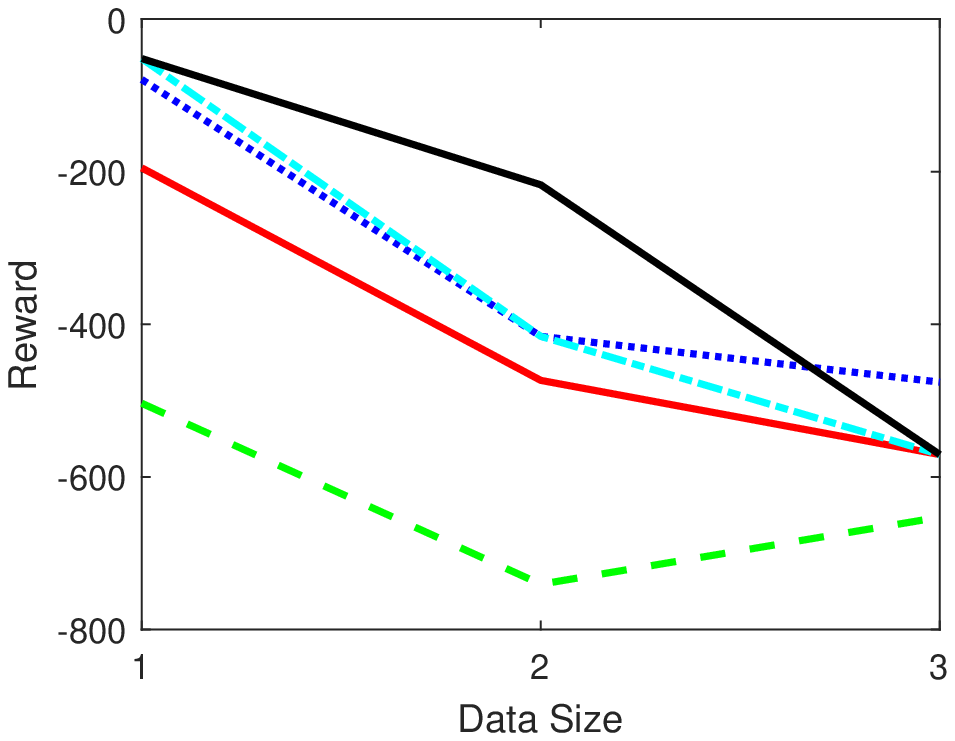}&
        \includegraphics[width=\figurewidth\textwidth]{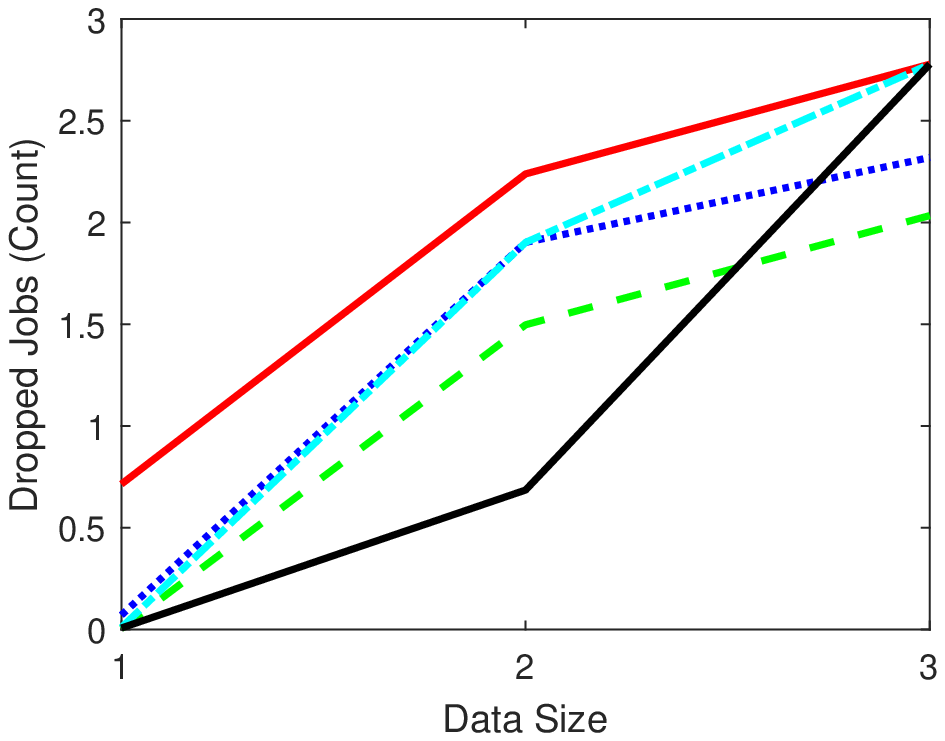}\\
        \includegraphics[width=\figurewidth\textwidth]{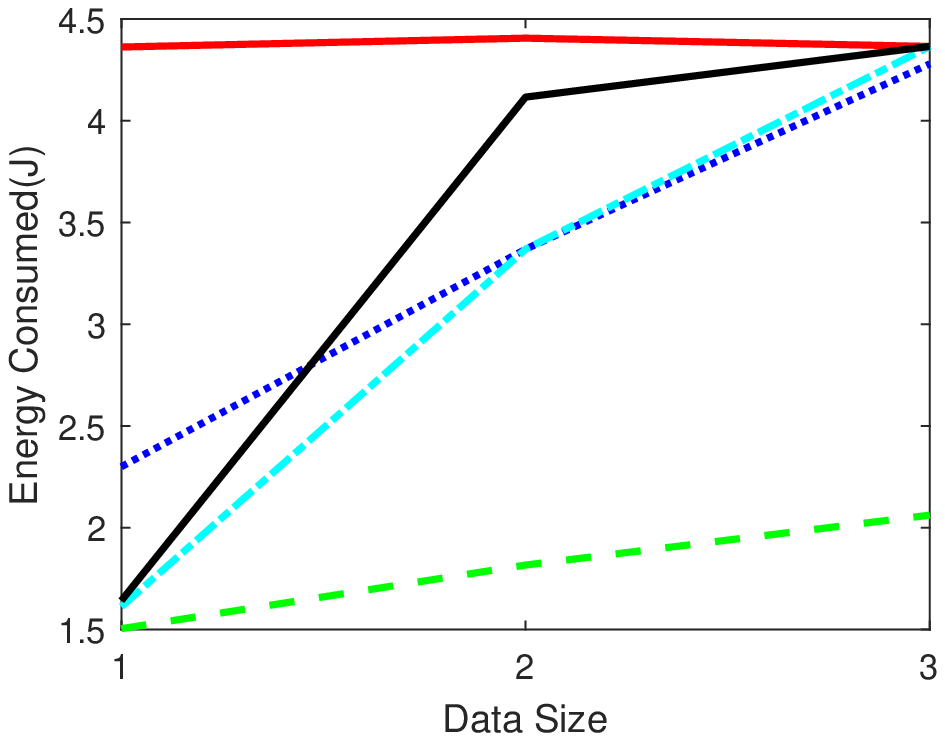}&\includegraphics[width=\figurewidth\textwidth]{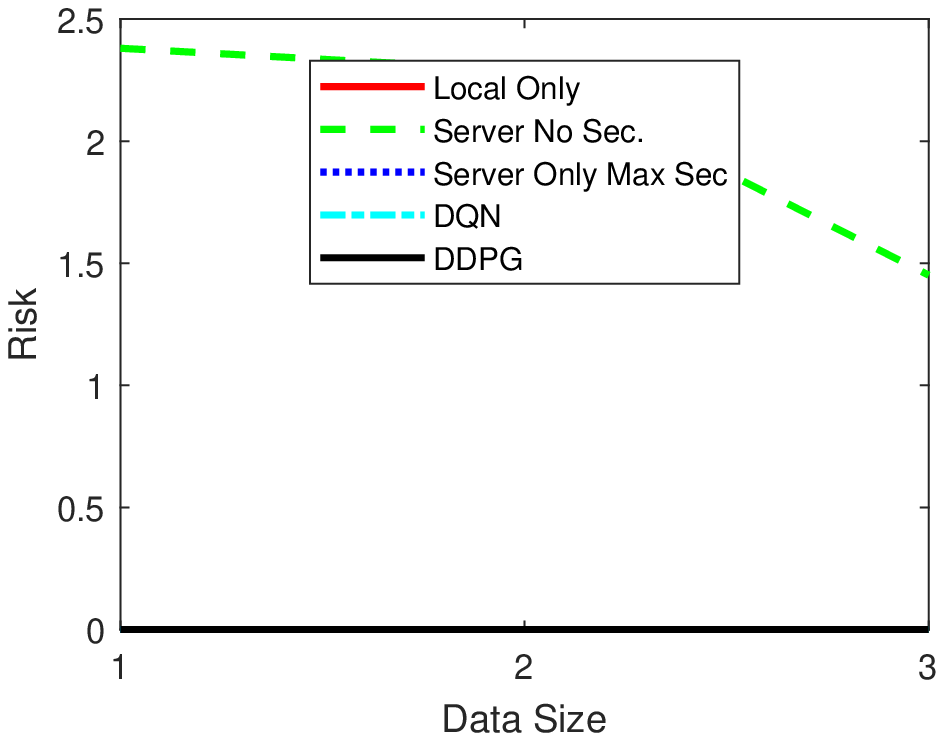}
    \end{tabular}
    \caption{Algorithm Performance for different data sizes (1, 2 or 3 times the default size). Clockwise from top left:  average overall score, average dropped jobs, lost jobs due to security, and energy consumed.}
    \label{fig:data_size}
\end{figure}

The size of data has multiple effects on the performance of the simulation. As the communication time is directly proportional to the data size, doubling the data size at least doubles the transmission time. In addition, our execution time is also proportional to the data size --see \cref{eqn_time}. So, the service rate of the satellite is halved when the data size is doubled. We present the results in \cref{fig:data_size}. We set the default data size of the algorithm to be 0.2 MB which could be the image data that the satellite is transmitting to the server for further computation. We see from the results that even under the same network and risk conditions, plenty of jobs are dropped when the data size is doubled. The \gls{lo} policy dropped on average $2.2$ jobs per episode followed closely by \gls{soms}, then by the \gls{sons}. Our \gls{ddpg} algorithm dropped the least with an average of only 0.5 jobs per episode. This improved performance came at a higher cost of energy. However, it still used less energy than the \gls{lo} with had the maximum energy consumption at all data sizes. However, when the data size is tripled, \gls{ddpg} algorithm dropped as many as the 2.7 jobs per episode which were the worst jointly with the local computing. In terms of overall cost, \gls{soms} was best at the triple data size.

\subsection{Satellite Formations}
{In \cref{subsec:const}, we described three different satellite formations namely leader/ follower, cluster and constellation. However, in previous sections, we experimented using the swarm/cluster formation only. The \gls{rtt} and the energy consumption can vary relative to the specific satellite formation \cite{radhakrishnanSurveyInterSatelliteCommunication2016}.  We capture the dynamics of this offloading scenario in our simulation by considering different transmission energy costs and varying the number of satellites in the communication (which in turn varies the communication data rate and time delay). For the satellites in the same orbit, the number of satellites would be limited which would mean the communication delay, as well as the energy cost, is lower. Similarly, in a cluster of satellites, the number of satellites transmitting simultaneously can be higher resulting in lower data rate and higher energy consumption. Finally, in a constellation, the number of satellites communicating would be still higher due to the larger area involved.}

\begin{figure}
    \centering
    \begin{tabular}{cc}
    \includegraphics[width=\figurewidth\textwidth]{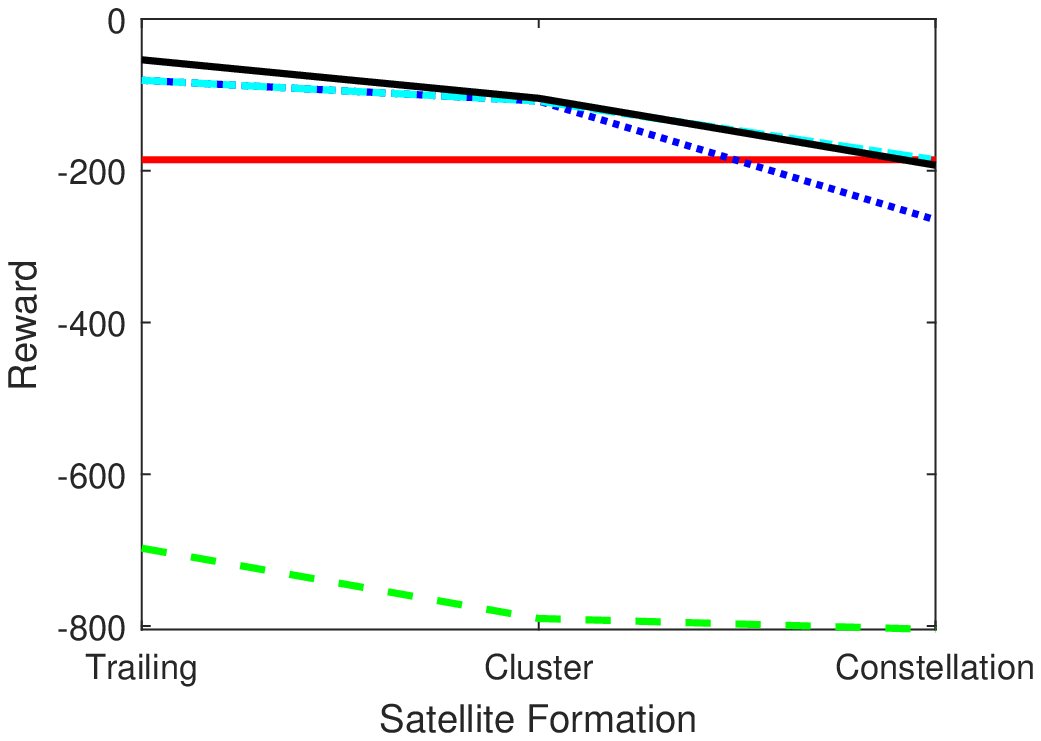}&
        \includegraphics[width=\figurewidth\textwidth]{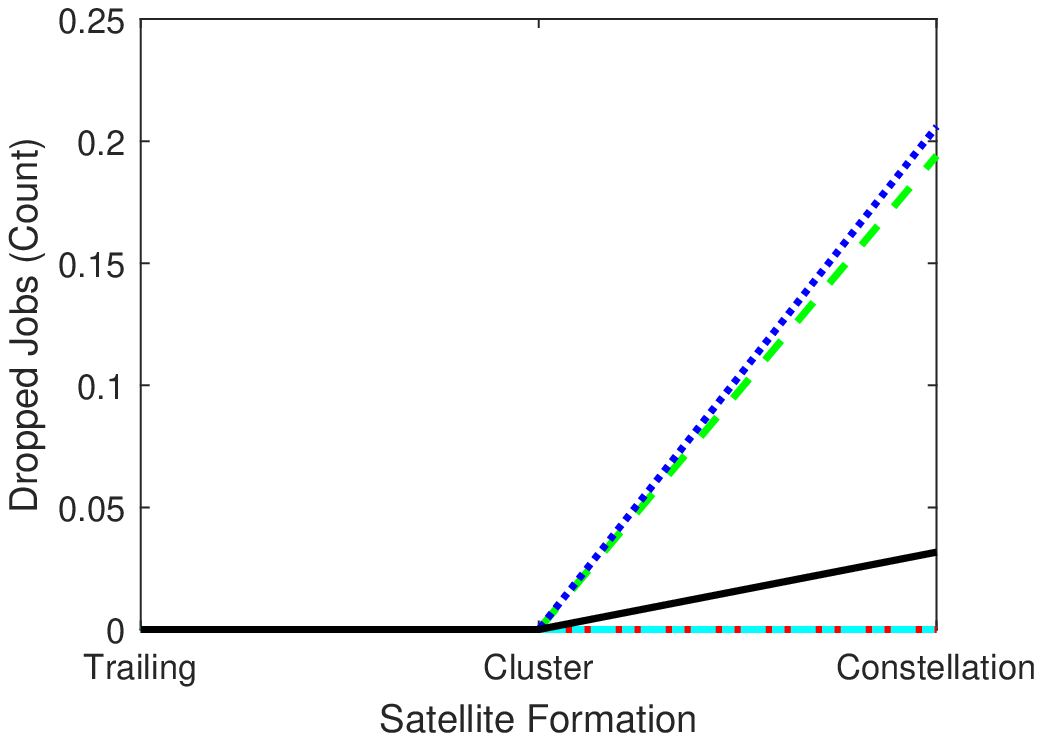}\\
        \includegraphics[width=\figurewidth\textwidth]{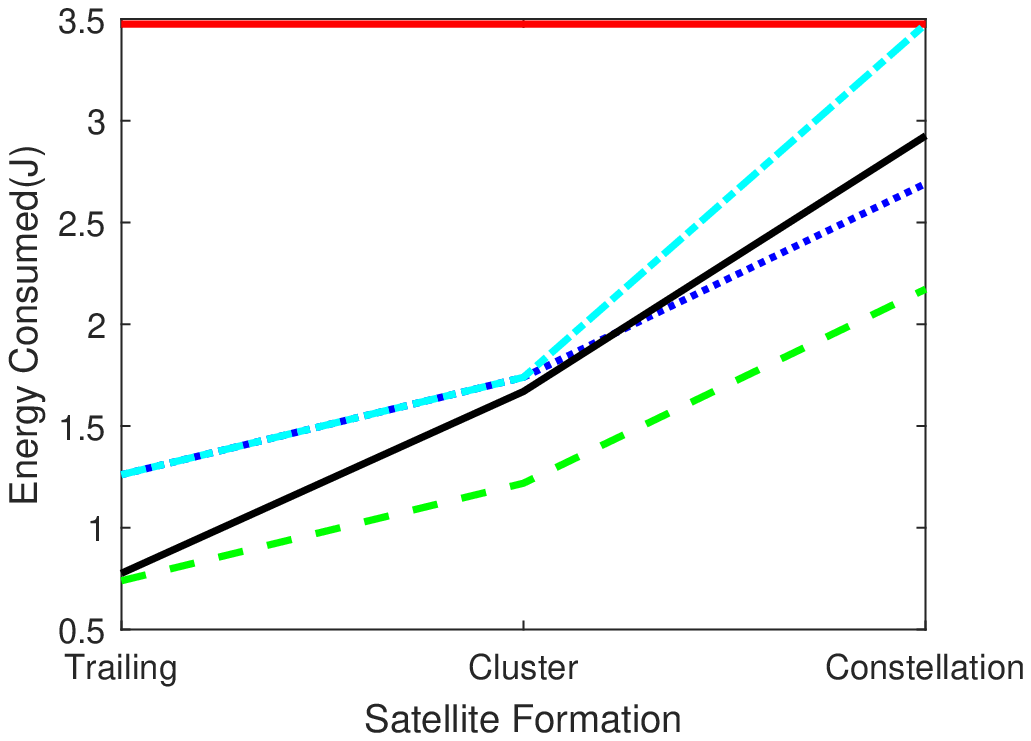}&\includegraphics[width=\figurewidth\textwidth]{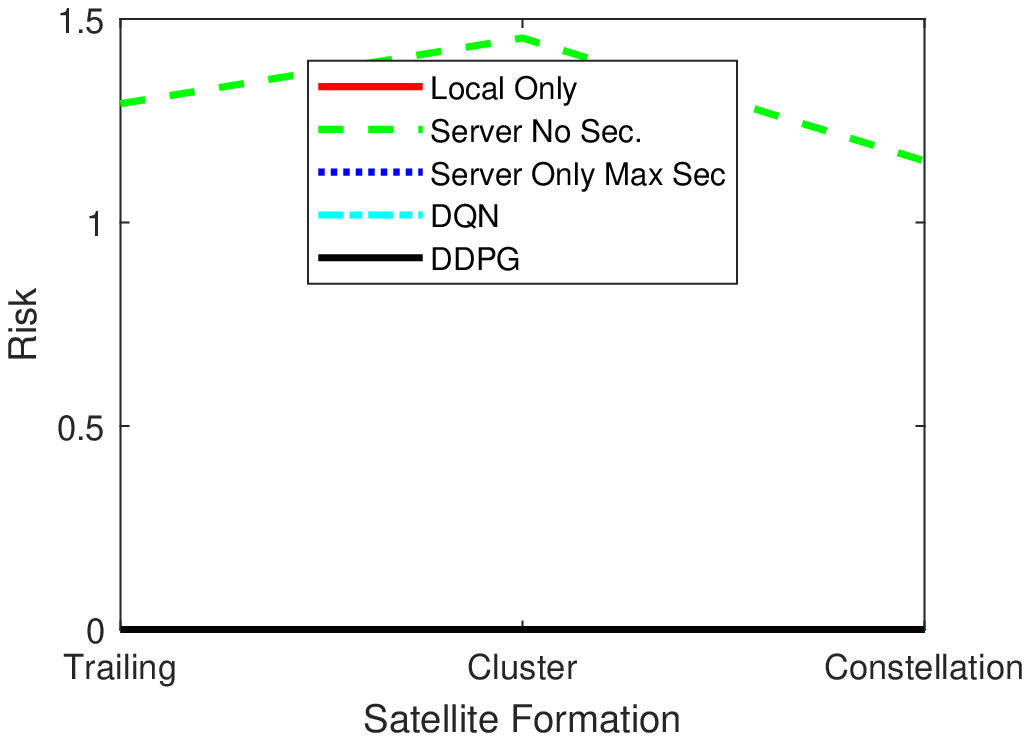}
    \end{tabular}
    \caption{Algorithm Performance for different satellite formations for low incoming job rate. Clockwise from top left:  average overall score, average dropped jobs, lost jobs due to security, and energy consumed.}
    \label{fig:const_load3}
\end{figure}

\begin{figure}
    \centering
    \begin{tabular}{cc}
    \includegraphics[width=\figurewidth\textwidth]{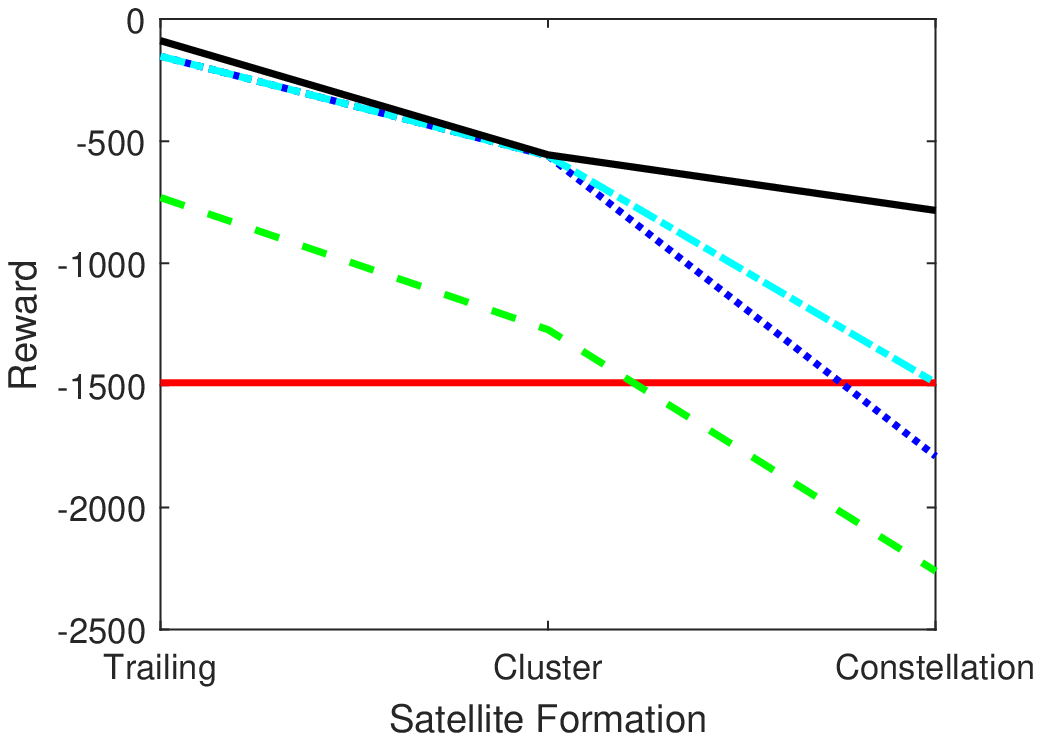}&
        \includegraphics[width=\figurewidth\textwidth]{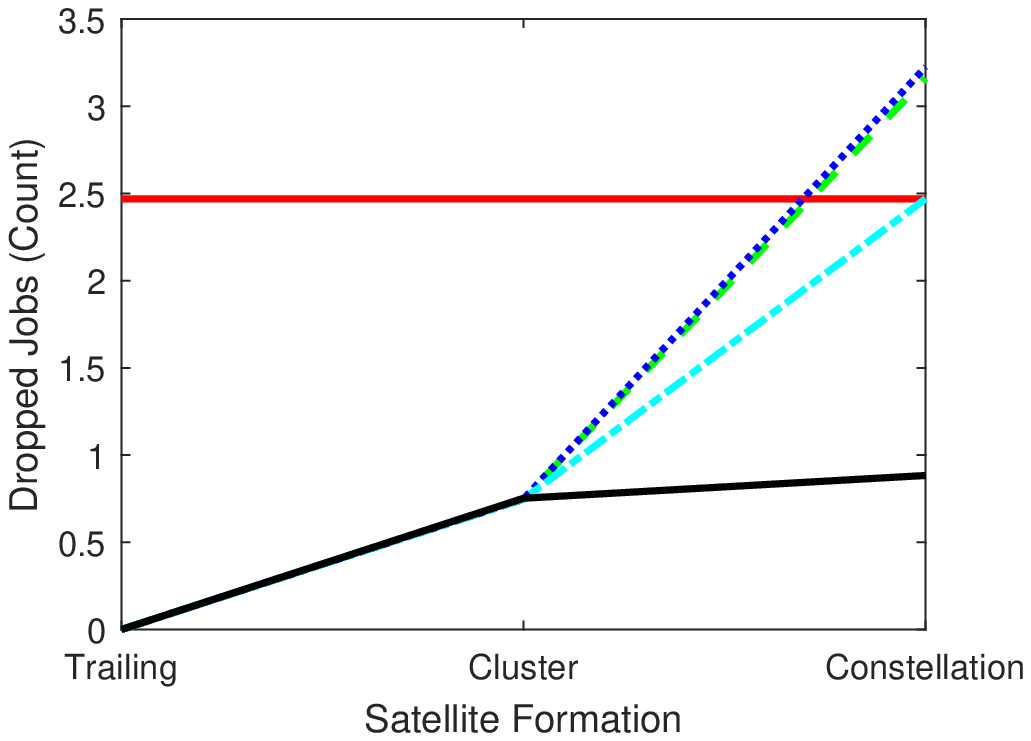}\\
        \includegraphics[width=\figurewidth\textwidth]{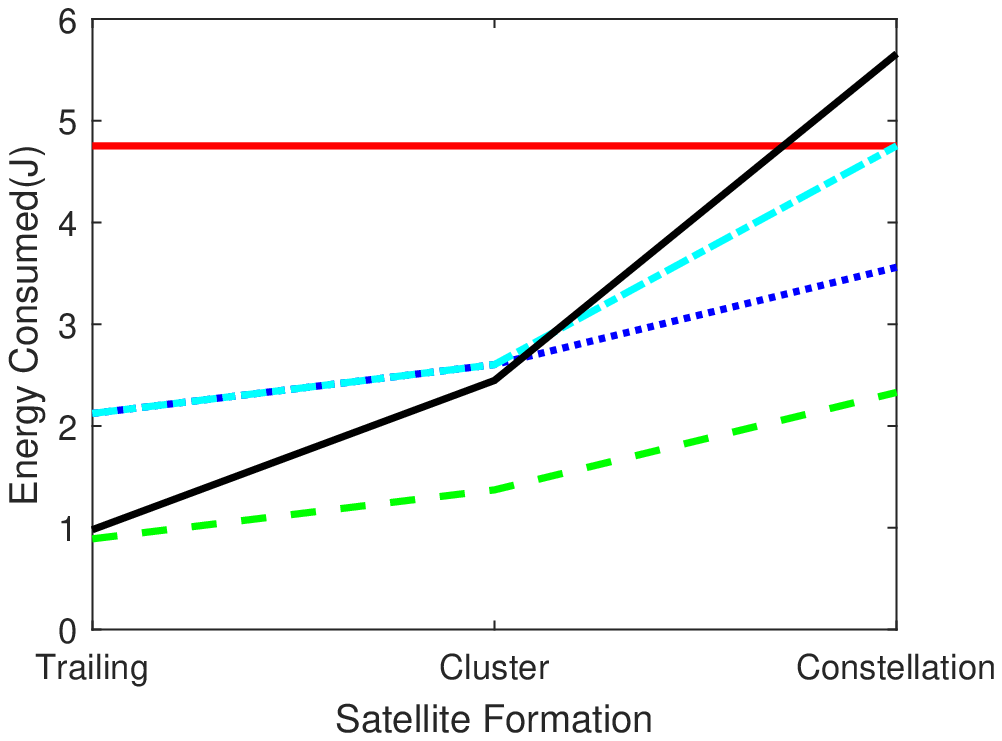}&\includegraphics[width=\figurewidth\textwidth]{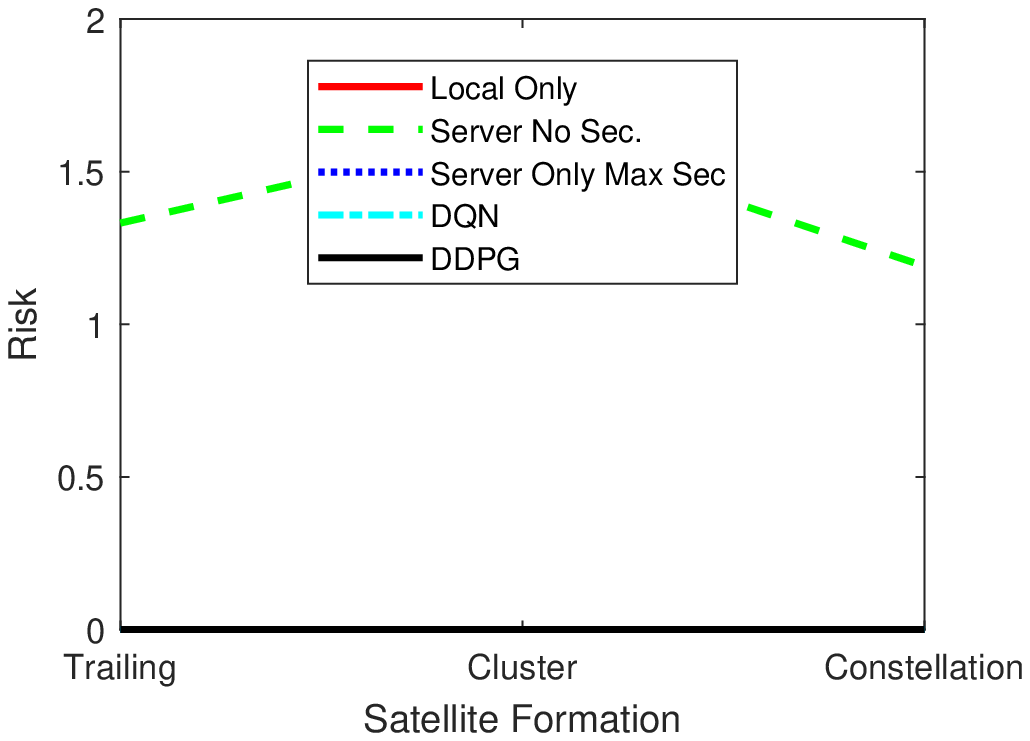}
    \end{tabular}
    \caption{Algorithm Performance for different satellite formations for high incoming job rate. Clockwise from top left:  average overall score, average dropped jobs, lost jobs due to security, and energy consumed.}
    \label{fig:const_load7}
\end{figure}

{\cref{fig:const_load3} and \cref{fig:const_load7} shows the results of the simulation for low and high incoming job rate cases respectively. In the low job rate case, none of the algorithms dropped any jobs for the trailing and cluster formation. For the constellation formation \gls{soms} and \gls{sons} policy both dropped approximately $0.2$ jobs per episode. Only \gls{sons} algorithms were subject to successful attacks as seen in the bottom right figure. However, from the bottom left figure, we see all the policies saved energy in comparison to the \gls{lo} policy. But the savings decreased as the formation changed from trailing to swarm and constellation. In the simulation setting when there were significantly more jobs present at the satellite (--see \cref{fig:const_load7}, we see the proposed algorithm \gls{ddpg} and \gls{dqn} was able to save energy as well as drop fewer jobs in comparison to the \gls{lo} case. The \gls{ddpg} was superior to the \gls{dqn} and others even in the constellation case where it dropped the least number of jobs.} 

{
From \cref{fig:const_load3} and \cref{fig:const_load7}, it is also evident that \gls{ddpg} is superior among all the policies. Also, it is evident that trailing formation is beneficial than local only, cluster, and the constellation formation for \gls{co}. This is because the communication channels are better than other formations. However, for all three formations, we simulated the same server capacity. As the formation grows larger in size, it may be possible to scale the server as well.}

\begin{table}[t]
    \centering
    \begin{tabular}{|l|c|c|}
    \hline
         \textbf{Formation}&\textbf{Power (Watt)}&\textbf{Maximum Satellites}  \\
         \hline
         Trailing&0.5&20\\
         Cluster&1&50\\
         Constellation&2&100\\
         \hline
    \end{tabular}
    \caption{Simulation parameters for different satellite formations}
    \label{tab:const_param}
\end{table}

\section{Conclusions and Future Work}\label{sec:con}
In this work, we studied wireless communications security for space applications. In addition, we applied a useful tool for future space applications called \acrlong{co}. We then solved the \gls{co} problem using a \gls{ddpg} algorithm which is a robust method for solving optimisation problems in real-time. Through extensive Monte-Carlo simulations, we show that our algorithm can increase the performance of space applications. The simulations also show that the added performance comes with increased energy consumption. We compared the proposed algorithm with not only static policies but with previously published \gls{dqn} method \cite{huangDeepReinforcementLearning2019}. In general, the experiments showed that the \gls{ddpg} is superior to \gls{dqn} based method in addition to static policies. {We also experimented on different satellite formations and show that the proposed algorithm is superior to the baselines. }However, in some settings, such as when the risk was very high, the \gls{dqn} agent performed better. Further study is necessary to understand the cause and to understand if multiple agents can be combined to develop an even better algorithm. For \gls{ns} system where nano-satellites could work in swarms and constellations with substantial autonomy, it is vital that satellites and rovers can trust each other and depend on each other. The algorithm presented in this paper can find applications in this new environment. In the future, we would like to incorporate further contexts such as the remaining energy resource of the satellites as well as the movement of satellites and authentication issues.

\ifCLASSOPTIONcompsoc
  \section*{Acknowledgments}
\else
  \section*{Acknowledgment}
\fi

This work is supported by grant EP/R026092 (FAIR-SPACE Hub) through UKRI under the Industry Strategic Challenge Fund  (ISCF)  for  Robotics  and  AI  Hubs  in  Extreme  and Hazardous Environments

\ifCLASSOPTIONcaptionsoff
  \newpage
\fi



\bibliographystyle{IEEEtran}
\bibliography{IEEEabrv,library,bibliography}
%

%








\end{document}

%% file: symbols.tex
\newglossaryentry{job_data}{name={\ensuremath{D}},description={Job data size}}
\newglossaryentry{t_deadline}{name={\ensuremath{\tau_d}},description={Maximum allowed time for completion of a job }}
\newglossaryentry{job_compute_per_bit}{name={\ensuremath{X}}, description={Number of instructions to execute a job per bit}}
\newglossaryentry{job_arrival_rate}{name={\ensuremath{\Lambda}},description={Job generation rate}}
\newglossaryentry{tm}{name={\ensuremath{\tau_m}}, description={Time taken by mobile device to execute the job}}
\newglossaryentry{Im}{name={\ensuremath{I_m}}, description={Number of instructions a device can execute per second}}
\newglossaryentry{Is}{name={\ensuremath{I_s}}, description={Number of instructions the server can execute per second}}
\newglossaryentry{Pc}{name={\ensuremath{P_m}}, description={Instantaneous power of mobile device while executing the job}}
\newglossaryentry{Em}{name={\ensuremath{E_{exec}}}, description={Total energy usage for executing the job}}
\newglossaryentry{Eoff}{name={\ensuremath{E_{off}}}, description={Total energy usage for offloading the job}}

\newglossaryentry{queue_servers}{name={\ensuremath{N_m}},description={Number of servers in queuing model equals number of cores in the satellite}}
\newglossaryentry{queue_servers_server}{name={\ensuremath{N_s}},description={Number of servers in queuing model equals number of cores in the space station}}
\newglossaryentry{util}{name={\ensuremath{u}},description={\gls{CPU} utilisation}}
\newglossaryentry{serv_rate}{name={\ensuremath{\mu}},description={Service rate of the node}, sort={service rate}}
\newglossaryentry{SD}{name={\ensuremath{SD}},description={Desired security level}}
\newglossaryentry{SP}{name={\ensuremath{SP}},description={Selected security level}}
\newglossaryentry{J}{name={\ensuremath{J}}, description={Maximum number of jobs in an interval}}
\newglossaryentry{comm_rate}{name={\ensuremath{R}}, description={Data Rate}}
\newglossaryentry{risk}{name={\ensuremath{r}},description={Risk}}
\newglossaryentry{state}{name={\ensuremath{\mathit{S}}},description={set of states for the \gls{mdp}}}
\newglossaryentry{action}{name={\ensuremath{\mathit{A}}},description={set of actions for the \gls{mdp}}}
\newglossaryentry{transition}{name={\ensuremath{\mathit{T}}},description={Transition function for the \gls{mdp}}}
\newglossaryentry{reward}{name={\ensuremath{\mathit{R}}},description={Reward function for the \gls{mdp}}}

%% file: glossary.tex
\newglossary[]{noshow}{}{}{Definitions}
\newglossaryentry{d_perc}{name={\ensuremath{D}},description={Percentage of security breach detected immediately}}
\newglossaryentry{test}{name={teset},description={This is a test}}

\newglossaryentry{app}
{
	name={\textit{app}},
	type=noshow,
	description={is an application program, generally developed for deployment in a Smartphone}
}
\newglossaryentry{hypothesis}
{
	name={\ensuremath{H}},
	description={is a hypothesis}
}
\newglossaryentry{M}{name={\ensuremath{M}},description={Number of rows in the image},sort={m}}
\newglossaryentry{N}{name={\ensuremath{N}},description={Number of columns in the image}}
\newglossaryentry{targets}{name={\ensuremath{p}},description={Number of targets in the camera}}
\newglossaryentry{max_targets}{name={\ensuremath{\gls{targets}_{max}}},description={Maximum number of targets, a camera can process in the allocated time}}
\newglossaryentry{off}{name={\textit{offloader}},type=noshow,description={The device that sends their job to cloud or neighbour for it to be executed and result sent back}}
\newglossaryentry{on}{name={\textit{onloader}},type=noshow, description={The device that takes on other's job for execution and sends result back after execution}}

\newglossaryentry{sensor_set}{name={\ensuremath{S}}, description={Total number of sensors deployed}}
\newglossaryentry{alg_set}{name={\ensuremath{\varLambda}},sort={L},description={Set of offloadable algorithms in sensor s}}

\newglossaryentry{feasible_set}{name={\ensuremath{\mathcal{F}}},description={Subset of \gls{sensor_set} that contains sensors feasible for offloading}}
\newglossaryentry{speed_factor}{name={\ensuremath{F}},description={Speed up factor of the \gls{on}}}

\newglossaryentry{constraint_set}{name={\ensuremath{Co}},description={Set of Constraints}}

\newglossaryentry{other_load}{name={\ensuremath{g}},description={Gaussian random variable}}

\newglossaryentry{off_ext_job}{name={\ensuremath{\gamma}}, sort={Offloadable external job},description={Incoming external job rate that are offloadable}}
\newglossaryentry{unoff_ext_job}{name={\ensuremath{\gamma_0}}, sort={Unoffloadable external job},description={Incoming external job rate that are not offloadable}}
\newglossaryentry{total_job}{name={\ensuremath{\lambda}},description={Total incoming job rate}, sort={total job}}

\newglossaryentry{qlength_job}{name={\ensuremath{L}},description={Number of jobs in the queue}}
\newglossaryentry{retrans_times}{name={\ensuremath{r}},description={Average retransmission times. When the channel is perfect its value is $0$}}
\newglossaryentry{arc_set}{name={\ensuremath{E}},description={Set of directed arcs between nodes}}
\newglossaryentry{adj_mat}{name={\ensuremath{A}},description={Adjacency matrix}}
\newglossaryentry{weight}{name={\ensuremath{\omega}},description={Weight factor}, sort={weight factor}}
\newglossaryentry{bat_level}{name={\ensuremath{BAT}},description={Remaining battery level}}
\newglossaryentry{proc_time}{name={proctime},description={Average processing time}}
\newglossaryentry{cost_fn}{name={\ensuremath{c}},description={Cost function of scheduling a job}}
\newglossaryentry{policy_mat}{name={\ensuremath{\mathbf{\hat X}}},description={Job schedule policy for all sensors}}
\newglossaryentry{policy_vec}{name={\ensuremath{ \vec{\hat x}}},description={Job schedule policy for one sensor}}
\newglossaryentry{mm1}{name={\ensuremath{M/M/1}}, type=noshow, description={A simple queue with one server, Poisson arrival, exponential service times, and infinite queue length}}
\newglossaryentry{util_q}{name={\ensuremath{\rho}},description={Utilisation of a queue},sort={utilisation of a queue}}
\newglossaryentry{nn_set}{name={\ensuremath{X_{NN}}},description={A set of nearest neighbours}}
\newglossaryentry{ref_set_no}{name={\ensuremath{N_r}},description={Number of Images in the reference set}}
\newglossaryentry{adj_const_set}{name={\ensuremath{\hat S}},description={Adjacency Constrained set}}
\newglossaryentry{sal_score}{name={\ensuremath{s_{knn}}},description={Saliency score}}
\newglossaryentry{dist_k}{name={\ensuremath{D_k}},description={Euclidean distance to the $k^{th}$ nearest neighbour}}
\newglossaryentry{dcolorsift}{name={\ensuremath{\vec{X}}},description={Dense Colour and \gls{sift} features (dColorSIFT)}}
\newglossaryentry{dcolorsift_row}{name={\ensuremath{\mathcal{T}}},description={Dense Colour and \gls{sift} features of a row}}

\newglossaryentry{radiopower}{name={\ensuremath{P}},description={Radio power (deciBel)}}
\newglossaryentry{distance}{name={\ensuremath{d}},description={Distance between transmitter and receiver}}
\newglossaryentry{antgain}{name={\ensuremath{G}},description={Gain of the antenna}}
\newglossaryentry{rate}{name={\ensuremath{R}},description={Data rate}}
\newglossaryentry{channelgain}{name={\ensuremath{H}},description={Channel gain}}
\newglossaryentry{backinterpower}{name={\ensuremath{\omega}},description={ denotes the background interference power}, sort={background interference power}}
\newglossaryentry{pktlength}{name={\ensuremath{\mathcal{D}}},description={Number of bits in a packet}}
\newglossaryentry{ber}{name={\ensuremath{\beta}}, sort={bit error rate}, description={\gls{BER} symbol}}
\newglossaryentry{errbit}{name={\ensuremath{{e}}},description={Maximum number of errors in a packet that can be corrected without resending the packet}}
\newglossaryentry{pathle}{name={\ensuremath{\eta}},description={Path loss exponent}, sort={path loss exponent}}
\newglossaryentry{conf_level}{name={\ensuremath{S_{conf}}},description={Encryption Security Level}}
\newglossaryentry{int_level}{name={\ensuremath{S_{int}}},description={Integrity Security Level}}

\newglossaryentry{q}{name={\ensuremath{Q}},description={Value of state action pair}}
\newglossaryentry{ns}{name={\textit{NewSpace}},description={NewSpace}}

%% file: acronyms.tex
\newacronym{DUT}{DUT}{Device Under Test}
\newacronym{OS}{OS}{Operating System}
\newacronym{sota}{SOTA}{State Of the Art}
\newacronym{mips}{MIPS}{Million Instructions Per Second}

\newacronym{BW}{BW}{\text{Network Bandwidth}}
\newacronym{AP}{AP}{Application Processor}
\newacronym{FOV}{FOV}{Field Of View}
\newacronym{CPU}{CPU}{Central Processing Unit}
\newacronym{DVFS}{DVFS}{Dynamic Voltage and Frequency Scaling}
\newacronym{GPU}{GPU}{Graphical Processing Unit}
\newacronym{SI}{SI}{Successful Identifications}
\newacronym{ES}{ES}{Efficiency Score}
\newacronym{OP}{OP}{Operations}
\newacronym{NO}{NO}{Non Offloading}
\newacronym{MEC}{MEC}{Minimal Energy Cost}
\newacronym{MBI}{MBI}{Minimal Battery Impact}
\newacronym{OOB}{OOB}{Offload Only if Busy}
\newacronym{IOT}{IOT}{Internet of Things}
\newacronym{VR}{VR}{Virtual Reality}
\newacronym{QOS}{QOS}{Quality of Service}
\newacronym{FPGA}{FPGA}{Field Programmable Gate Array}
\newacronym{cots}{COTS}{Commercial Off-the-Shelf}
\newacronym{FCFS}{FCFS}{First Come First Service}
\newacronym{BER}{BER}{Bit Error Rate}
\newacronym{PDR}{PDR}{Packet Delivery Rate}
\newacronym{O}{O}{Oracle}
\newacronym{PC}{PC}{Proactive Centralised}
\newacronym{PD}{PD}{Proactive Distributed}
\newacronym{RD}{RD}{Reactive Distributed}
\newacronym{NSI}{NSI}{Node State Information}
\newacronym{MCC}{MCC}{Mobile Cloud Computing}
\newacronym{RFH}{RFH}{Request For Help}
\newacronym{HSV}{HSV}{Hue Saturation Value}
\newacronym{PCA}{PCA}{Principal	Component Analysis}
\newacronym{pdf}{pdf}{probability distribution function}
\newacronym{dv}{dv}{\textit{decision vector}}
\newacronym{RWP}{RWP}{Random Waypoint Model}
\newacronym{RTT}{RTT}{Round Trip Time}
\newacronym{MDP}{MDP}{Markov Decision Process}
\newacronym{PARSEC}{PARSEC}{Princeton Application Repository for Shared-Memory Computers}
\newacronym{XML}{XML}{Extensible Markup Language}

\newacronym{NN}{NN}{Nearest Neighbour}
\newacronym{RAM}{RAM}{Random Access Memory}
\newacronym{ADK}{ADK}{Android Development Kit}
\newacronym{LDML}{LDML}{Logistic Discriminant Metric Learning}
\newacronym{CCTV}{CCTV}{Closed Circuit Television}

\newacronym{1GSS}{1GSS}{first generation surveillance system}
\newacronym{2GSS}{2GSS}{second generation surveillance system}
\newacronym{3GSS}{3GSS}{third generation surveillance system}
\newacronym{4GSS}{4GSS}{fourth generation surveillance system}
\newacronym{dstl}{DSTL}{Defence Science and Technology Laboratory}
\newacronym{JVM}{JVM}{Java Virtual Machine}
\newacronym{FSM}{FSM}{Finite State Machine}
\newacronym{FFT}{FFT}{Fast Fourier Transform}
\newacronym{MIPSA}{MIPSA}{Modular Integrated Passenger Surveillance Architecture}

\newacronym{fob}{FOB}{forward operating base}

\newacronym{apm}{APM}{Active Power Management}
\newacronym{manet}{MANET}{Mobile Ad-hoc Network}
\newacronym{iot}{IOT}{Internet Of Thing}
\newacronym{uav}{UAV}{Unmanned Aerial Vehicle}
\newacronym{fcfs}{FCFS}{First Come First Service}
\newacronym{mc}{MC}{Markov Chain}
\newacronym{ic}{IC}{Integrated Circuit}

\newacronym{CV}{CV}{Computer Vision}
\newacronym{RGB}{RGB}{Red Green Blue}
\newacronym{hog}{HOG}{Histogram of Gradients}
\newacronym{LBP}{LBP}{Local Binary Pattern}
\newacronym{sift}{SIFT}{Scale Invariant Feature Transform}
\newacronym{WHSV}{WHSV}{Weighted Color Histograms}
\newacronym{MSCR}{MSCR}{Maximally Stable Color Region}
\newacronym{RHSP}{RHSP}{Recurrent High-Structured Patches }
\newacronym{MOG}{MOG}{Mixture of Gaussians}
\newacronym{svm}{SVM}{Support Vector Machine}
\newacronym{knn}{k-NN}{k-Nearest Neighbour}
\newacronym{nn}{NN}{Neural Network}

\newacronym{PRID}{PRID}{Person Re-identification}
\newacronym{LAFT}{LAFT}{Locally Aligned Feature Transform}
\newacronym{ITML}{ITML}{Information Theoretic Metric Learning}
\newacronym{KISSME}{KISSME}{Keep It Simple and Straightforward MEtric}
\newacronym{SDALF}{SDALF}{Symmetry-Driven Accumulation of Local Features}
\newacronym{bicov}{BiCov}{Bio-inspired Covariance based features}
\newacronym{SDC}{SDC}{Saliency guided Dense Correspondence}
\newacronym{mars}{MARS}{Motion Analysis and Re-identification Set}
\newacronym{VIPeR}{VIPeR}{Viewpoint Invariant Pedestrian Recognition}
\newacronym{ROC}{ROC}{Receiver Operating Characteristics}
\newacronym{cmc}{CMC}{Cumulative Matching Characteristics}
\newacronym{DNN}{DNN}{Deep Neural Network}
\newacronym{CNN}{CNN}{Convolutional Neural Network}
\newacronym{LSTM}{LSTM}{Long Short Term Memory}
\newacronym{GAN}{GAN}{Generative Adversarial Networks}
\newacronym{ReLU}{ReLU}{Rectified Linear Unit}
\newacronym{FPS}{FPS}{Frames Per Second}

\newacronym{rtt}{RTT}{Round Trip Time}
\newacronym{SNR}{SNR}{Signal to Noise Ratio}
\newacronym{sinr}{SINR}{Signal to Interference and Noise Ratio}
\newacronym{rssi}{RSSI}{Received Signal Strength Indication}
\newacronym{csma}{CSMA}{Carrier Sense Multiple Access}
\newacronym{wcdma}{W-CDMA}{Wideband-Code Division Multiple Access}
\newacronym{FACH}{FACH}{Forward Access Channel}
\newacronym{DCH}{DCH}{Dedicated Channel}
\newacronym{UE}{UE}{User Equipment}
\newacronym{RNC}{RNC}{Radio Network Controller}
\newacronym{RRC}{RRC}{Radio Resource Control}
\newacronym{HSUPA}{HSUPA}{High-Speed Uplink Packet Access}
\newacronym{HSDPA}{HSDPA}{High-Speed Downlink Packet Access}
\newacronym{dos}{DOS}{Denial Of Service}
\newacronym{dsr}{DSR}{Dynamic Source Routing}
\newacronym{aodv}{AODV}{Ad-hoc On-demand Distance Vector}
\newacronym{tora}{TORA}{Temporarily Ordered Routing Algorithm}
\newacronym{dsdv}{DSDV}{Destination-Sequenced Distance-Vector}
\newacronym{olsr}{OLSR}{Optimised Link State Routing}
\newacronym{zrp}{ZRP}{Zonal Routing Protocol}
\newacronym{cluspow}{CLUSTERPOW}{long}
\newacronym{hppp}{HPPP}{Homogeneous Poisson Point Process}
\newacronym{ppp}{PPP}{Poisson Point Process}
\newacronym{VANET}{VANET}{Vehicle Ad-hoc Network}
\newacronym{co}{CO}{Computation Offloading}
\newacronym{mcc}{MCC}{Mobile Cloud Computing}
\newacronym{mec}{MEC}{Mobile Edge Computing}
\newacronym{geo}{GEO}{geo stationary}
\newacronym{leo}{LEO}{Low Earth Orbit}
\newacronym{meo}{mEO}{Medium Earth Orbit}
\newacronym{gps}{GPS}{Global Positioning System}

\newacronym{rl}{RL}{Reinforcement Learning}
\newacronym{dqn}{DQN}{Deep Q-Network}
\newacronym{ddqn}{DDQN}{Double \gls{dqn}}
\newacronym{ddpg}{DDPG}{Deep Deterministic Policy Gradient}

\newacronym{cia}{CIA}{Confidentiality Integrity Availabilty}
\newacronym{mdp}{MDP}{Markov Decision Process}
\newacronym{ue}{UE}{User Equipment}

\newacronym{iost}{IOST}{Internet of Space Things}
\newacronym{iov}{IOV}{Internet of Vehicles}
\newacronym{ra}{RA}{Reference Architecture}
\newacronym{nasa}{NASA}{National Aeronautics and Space Administration}
\newacronym{sdn}{SDN}{Software Defined Networking}
\newacronym{sme}{SME}{Small and Medium Enterprise}
\newacronym{dtn}{DTN}{Delay Tolerant Network}

\newacronym{aes}{AES}{Advanced Encryption Standard}
\newacronym{qos}{QOS}{Quality of Service}
\newacronym{vnf}{VNF}{Virtualised Network Function}

\newacronym{lo}{LO}{Local Only}
\newacronym{sons}{SONS}{Server Only No Security}
\newacronym{soms}{SOMS}{Server Only Maximum Security}